\documentclass[numberedappendix]{openjournal}

\usepackage{etoolbox}
\usepackage{lipsum}
\usepackage{graphicx}
\usepackage{subcaption}
\usepackage{amsmath}

\usepackage{xcolor}
\usepackage{textgreek}
\usepackage[utf8]{inputenc}
\usepackage[english]{babel}

\definecolor{linkcolor}{rgb}{0.0,0.3,0.5}
\usepackage{tensind}
\tensordelimiter{?}
\DeclareGraphicsExtensions{.bmp,.png,.jpg,.pdf}
\usepackage{verbatim}
\usepackage[normalem]{ulem}
\usepackage{orcidlink}
\usepackage{soul}

\newcommand{\vk}{\vec{k}}

\definecolor{dragonsbreath}{HTML}{cc0044}
\usepackage{hyperref}
\hypersetup{
    unicode, 
    colorlinks=true,
    linkcolor=linkcolor,
    citecolor=linkcolor,
    filecolor=linkcolor,
    urlcolor=linkcolor,
}
\graphicspath{ {./figs/} }
\begin{document}
\title{One with HI: Modelling HI Intensity Mapping one-point statistics\\including systematics}

\author{Bernhard Vos-Gines\orcidlink{0000-0002-1803-1169}}
\author{Cora Uhlemann\orcidlink{0000-0001-7831-1579}}
\affiliation{Fakultät für Physik, Universität Bielefeld, Postfach 100131, 33501 Bielefeld, Germany}
\email{bvosgines@physik.uni-bielefeld.de}


\begin{abstract}
    Neutral hydrogen (HI) traces the dark matter distribution of the Universe. Upcoming surveys such as the Square Kilometre Array Observatory (SKAO) will trace neutral hydrogen up to $z \leq 6$ using several detection techniques including Intensity Mapping, which offers a unique window to explore the post-reionization Universe. Beyond two-point statistics promise to extract additional non-Gaussian information but require an accurate modelling of observational systematics such as foregrounds and the telescope beam. This work develops a theoretical model for the HI one-point probability density function (PDF) in spherical cells based on large-deviation statistics and spherical collapse for dark matter along with a nonlinear tracer bias and stochasticity parameterisation. It incorporates foreground removal and telescope beam effects that are validated against high-resolution simulations. We show that, despite these observational systematics, the HI PDF is able to capture additional non-Gaussian information from HI intensity maps compared to the power spectrum and can thus tighten constraints on cosmological parameters, breaking the degeneracy between the linear bias and the clustering amplitude.  
    
\end{abstract}


\maketitle

\section{Introduction}
\label{sec:intro}
To take maximum profit of the increase of cosmological information already provided by current Stage IV surveys such as DESI \citep[DESI,][]{desicollaboration2016,DESI2022} and Euclid \citep{laureijs2011euclid} and expected from future surveys such as the Vera C. Rubin Observatory \citep[LSST,][]{Ivezic_2019}, WFIRST \citep{akeson2019} and the Square Kilometre Array Observatory (SKAO) \citep{Bacon_2020}, cosmologists focus their attention on summary statistics that are able to capture non-Gaussian information. Harnessing the power of beyond two-point statistics requires modelling of the cosmological signal and the impact of systematics, which we provide here for the one-point probability density function of the HI intensity on mildly nonlinear scales.

SKAO is a radio survey that will map neutral hydrogen (HI) on an unprecedented volume of the post-reionization Universe $z \leq 6$. HI is detected via its 21-cm hyperfine transition between parallel and antiparallel spin states. Although it is a rare transition, the vast amount of HI gas in the Universe allows HI detection with high signal-to-noise. In the post-reionization Universe, HI can be found in self-shielded environments, associated with galaxies.   

One of the SKAO main programs to observe neutral hydrogen is the HI Intensity Mapping (IM) wide Survey of 20000 $\rm{deg}^2$ for $0.35 < z < 3$. HI Intensity Mapping \citep{Bharadwaj_2001,Battye_2004,chang_2008} measures the integrated brightness temperature fluctuations across the sky as a function of redshift. Thus, IM can map vast volumes very efficiently without the need of resolving individual galaxies, and it integrates all the signal without the detection thresholds required by spectroscopic surveys. Furthermore, this technique has the potential to map mostly unexplored regions of the Universe, where high-redshift spectroscopy becomes limited.  In this work we focus on the auto-correlation data of the telescope array, or single-dish Intensity Mapping \citep{Battye2013} instead of its interferometric mode, which gives us access to the largest scales. 

To extract reliable cosmological information from the SKAO Intensity Mapping survey we need to take into account several observational systematics. The beam effect of the telescope antennas introduces an angular smoothing in the signal that increases with redshift and is inversely proportional to the antenna's diameter. Nevertheless, the spectroscopic resolution is excellent which allows probing small line-of-sight scales \citep{Villaescusa_2017}. 

Another important challenge is the foregrounds from astrophysical sources dominated by galactic synchrotron emission. Photons emitted by electrons accelerated by Galactic magnetic fields share the same frequency as our redshifted 21-cm line in our observer frame, making it challenging to disentangle the cosmological 21-cm signal from the Galactic foreground emission. Foregrounds are usually several orders of magnitude brighter than the cosmological signal but they are smooth in frequency, which makes them in principle distinguishable from the fluctuating HI signal. However, the removal of foregrounds damps the largest radial scales, that is, the small-$k_{\parallel}$ modes \citep{Cunnington_2021}.

Other systematics like the shot noise should be sub-dominant for Intensity Mapping surveys, but we should consider the thermal noise of the instrument, which depends on the system temperature, the observed area and observed time \citep{Battye2013}. We model both noise sources in this work for Intensity Maps of voxel size $V_{\rm cell} = L_{\rm cell}^3 = 1$ $({\rm Mpc}/h)^3$.  

The power spectrum, $P(k)$, has been extensively used to extract cosmological information from a wide variety of surveys, including recently HI Intensity Mapping with the first detection of the HI Intensity Mapping Power spectrum from interferometric observations of the MeerKAT radio telescope, a precursor of SKA \citep{paul2023}. Nevertheless, in the recent years, beyond two-point statistics are gaining a lot of attention in the literature. Those statistics aim to extract non-Gaussian information from the field that is missing in two-point statistics \citep{Majumdar2026}.  Beyond N-point correlation functions such as the bispectrum \citep{Majumdar_2018} or trispectrum \citep{Cooray_2008}, one-point statistics such as voxel intensity distributions \citep{Breysse_2017} or topological statistics  such as Minkowski functionals \citep{Gleser_2006,Spina_2021} or Betti numbers \citep{Giri_2021} have been considered.
The one-point probability density function (PDF) of HI intensities can harvest non-Gaussian information and be predicted in the mildly nonlinear regime, making it an ideal complement to the power spectrum. 
The analytical matter PDF shape was first theorized by \citep{bernardeau1994} based on tree-level perturbation theory. Further improvements using Large Deviation Theory and the Spherical Collapse were introduced to improve the agreement with N-body simulations spanning a wide range of densities \citep{Bernardeau_2014,Uhlemann_2016}. The sensitivity of the PDF on cosmological parameters has been tested \citep{Friedrich_2021,Uhlemann_2018b,Uhlemann_2020} and applied to biased tracers such as halos and galaxies \citep{Uhlemann_2018a,Friedrich_2021a,Gould2025}. The neutral hydrogen PDF has been previously related to the matter PDF using an empirical abundance matching bias model from simulations \citep{Leicht2019}. PDFs of densities and intensities quantify the properties of different density environments at a given smoothing scale and are closely related to the computationally efficient $k$-Nearest Neighbour statistics \citep[see e.g.][]{BanerjeeAbel2021,BanerjeeKokron2022,Coulton2024}, which organise this information in terms of the distribution of distances of nearest neighbours.

In this work we model the HI PDF including a theoretical bias and stochasticity model and an incorporation of observational systematics. We will show how in this setup, when the HI PDF is included in the analysis, it provides complementary information to the power spectrum, improving the constraints in the $\Omega_{\rm M}-\sigma_8$ plane. Together with the power spectrum, we can use it to lift the degeneracy between the linear bias $b_1$ and the amplitude of fluctuations $\sigma_8$.

This paper is structured as follows: in \autoref{sec: data} we present the simulation data we use in this work. In \autoref{sec: observational_systematics} we introduce the most important observational systematics we need to model an HI Intensity mapping survey: the telescope beam, foreground-removal and thermal noise. We provide theoretical models for our summary statistics, the HI power spectrum and one-point PDF in \autoref{sec:model_test_summary_statistics} and we test them to summary statistics which are simulation-based. In \autoref{section: cosmological_parameter_estimation} we illustrate our summary statistic covariances, determine their sensitivity to cosmological parameters and forecast constraints for several cosmological parameters. Finally, in \autoref{sec:concl} we summarize our results.

\section{Simulation data}
\label{sec: data}

In this section we describe the steps we follow to compute our simulated mock HI intensity maps, used to validate a theoretical model for the power spectrum and the one-point probability density function in further sections. The pipeline includes the usage of UNIT simulations \citep[][]{Chuang_2019} to obtain the matter and halo catalogues, a semianalytical model of galaxy evolution (SAGE) to obtain galaxy properties from the haloes (and in particular the cold gas mass distribution), an HI prescription as a function of the cold gas mass and a Nearest Grid Point (NGP) assignment of HI masses to obtain the HI intensity map. Finally, we convert our HI mass Intensity maps to HI temperature maps.

\subsection{UNITsims haloes}

In this work, we make use of the UNIT cosmological simulations \citep[UNITsims,][]{Chuang_2019}, a suite of four high-resolution dark-matter-only N-body  simulations (\textsc{UNIT1},\textsc{UNIT1InvPhase},\textsc{UNIT2} and \textsc{UNIT2InvPhase}). All simulations contain $4096^3$ particles of mass $m_p = 1.25 \times 10^{9}$ $h^{-1} M_\odot $, have a volume of $V = 1$ $h^{-3} \rm{Gpc}^{3}$ and share the same cosmology based on Planck 2015 results: $\Omega_{\rm M} = 1 - \Omega_\Lambda = 0.3089$, $h = 0.6774$, $n_s = 0.9667$ and $\sigma_8 = 0.8147$. Unless stated otherwise, all simulation-based results use the four simulations described above.

Dark matter particles were evolved to $z=0$ using the
TreePM code \textsc{L-Gadget}, a version of \textsc{Gadget2} \citep[][]{springel}, with initial conditions generated at $z_{\rm ini} = 99$ using the public code \textsc{FastPM} \citep[][]{fastpm}. Haloes were identified using the phase-space halo finder \textsc{Rockstar} \citep[][]{behroozi}. A key feature of the UNIT suite is the use of paired-and-fixed initial conditions \citep[][]{Angulo2016}, which significantly reduce sample variance on large scales by fixing the amplitude of Fourier modes and pairing realizations with opposite initial phases. Although this approach reduces the scatter in two-point statistics, the scatter in other clustering statistics—such as counts-in-cells, which we use in this work—remains the same as in simulations without fixed and paired initial conditions \citep[][]{Villaescusa_Navarro_2018}. Moreover, this technique does not introduce any measurable bias in one-, two-, or three-point statistics \citep[][]{Angulo2016,Villaescusa_Navarro_2018,Chuang_2019}.

\subsection{HI-Halo connection}

The Semi-Analytic Galaxy Evolution model \citep[SAGE,][]{sage} was used to assign galaxy properties to the dark matter haloes from the UNITsims \citep[][]{Knebe_2022}. Semi-analytical models assign gas to haloes at very early times and evolve galaxy properties along the merger trees. The physical processes relevant for the formation and evolution of galaxies are encapsulated in several differential equations such as star formation, gas cooling, supernova feedback etc. SAGE has eight free parameters which are calibrated to reproduce the $z=0$ stellar mass function. 

We are particularly interested in the cold gas mass distribution $M_{\rm CG}$ output, and we will only consider objects with cold gas mass $M_{\rm CG} > 10^{2}$ $h^{-1}M_{\odot}$, stellar mass $M_{\rm \ast} > 10^{9}$ $h^{-1}M_{\odot}$ and halo mass $M_{h} > 10^{10.5}$ $h^{-1}M_{\odot}$ following \citet{Knebe_2022}. This gives us a HI galaxy catalogue with a number density of $\bar{n}_{\rm HI} = 5.9 \times 10^{-2} \left(h/\rm{Mpc} \right)^3$. The cold gas mass is then converted to neutral hydrogen mass as 
\begin{equation}
    M_{\rm HI} = f_{\rm H} \left(1-\frac{R_{\rm mol}}{R_{\rm mol}+1}\right) M_{\rm CG} \, ,
\end{equation}
where $f_{\rm H}$ is the hydrogen fraction and $R_{\rm mol}=0.4$ is the molecular-to-atomic ratio which is considered constant \citep{Zoldan2016,Zwaan2005}.

\subsection{HI Intensity maps}

We sum the HI masses of our galaxies in cubic voxels of side length $L_{\rm cell} = 1$ ${\rm Mpc}/h$ using the nearest-grid-point (NGP) assignment scheme and then we compute the HI mass overdensity. 
For intensity mapping, the angular resolution is typically worse than the transverse voxel scale even at relatively low redshifts due to the telescope beam effect (see section \ref{sec: telescope_beam}). In contrast, the radial resolution is typically higher thanks to the very precise redshift measurements of the HI line. In this work we use cubic voxels due to the finite resolution of numerical simulations. The observable in HI Intensity Mapping surveys is the difference in the brightness temperature $\Delta T_{\rm HI} = \bar{T}_b \delta_{\rm HI}$, where $\bar{T}_b$ is the mean brightness temperature (see \citet{Battye2013} for details)
\begin{equation}
    \bar{T}_\text{b}(z) = 190 \, \frac{H_0 \left(1+z\right)^2}{H(z)}\Omega_{\rm HI}(z) \,h \, ,
\end{equation}
where $\Omega_{\rm HI}(z) $ is the fraction of neutral hydrogen in the Universe at redshift $z$. 

\begin{figure}[htbp]
    \centering
    \begin{subfigure}[t]{0.48\textwidth}
        \centering
        \includegraphics[width=\textwidth]{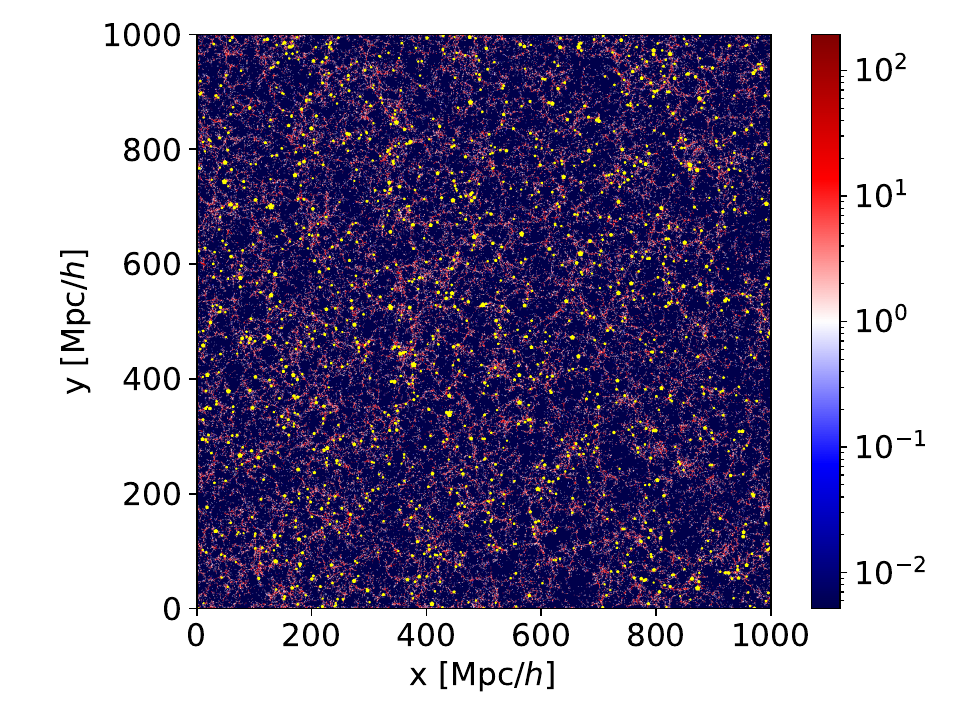}
        \caption{HI map without systematics}
        \label{fig:unit_no_sys}
    \end{subfigure}
    \hfill
    \begin{subfigure}[t]{0.48\textwidth}
        \centering
        \includegraphics[width=\textwidth]{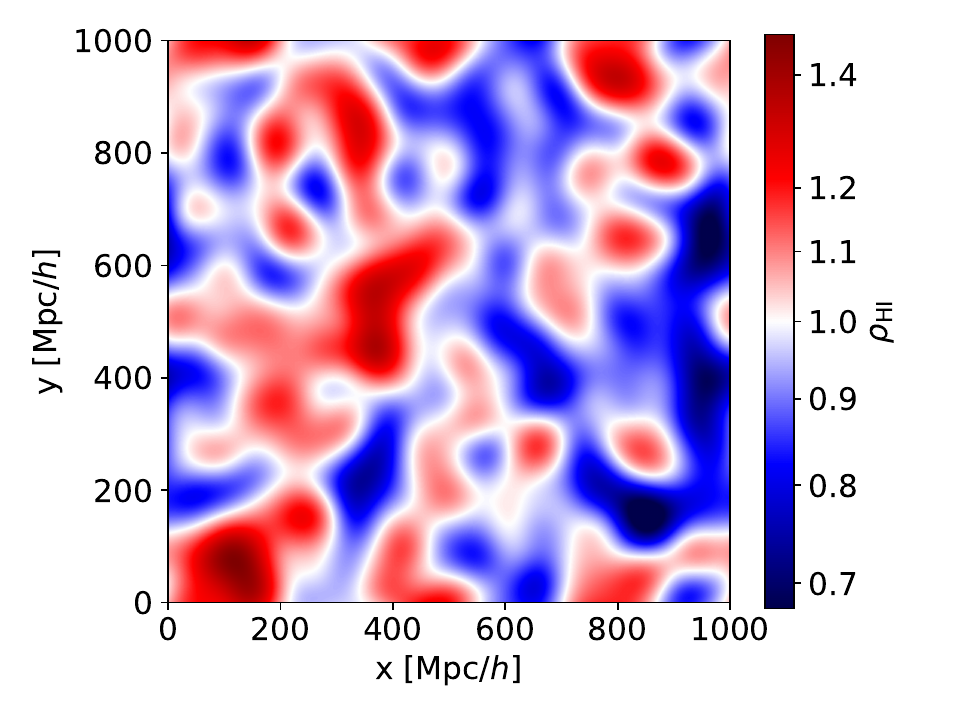}
        \caption{HI map with systematics}
        \label{fig:unit_sys}
    \end{subfigure}
    \caption{\textbf{Left: } We represent $\rho_{\rm HI} = 1+\Delta T_{\rm HI}/\bar{T}_b$ on a $5$ ${\rm Mpc}/h$ slice of the HI intensity map without observational systematics obtained from the simulation box \textsc{UNIT1} at redshift $z = 1.321$. Overlaid on this, we show 0.5 percent of the HI galaxies as yellow stars with a size proportional to their HI mass. \textbf{Right: } Same slice from an HI Intensity map with systematics included. }
    \label{fig:SIM}
\end{figure}

In this work we use the snapshot at redshift $z=1.321$, where $\Omega_{\rm HI}(z=1.321)= 3.8 \times 10^{-4}$ and $\bar{T}_\text{b}(z=1.321) = 1.82 \times 10^{-1}$ mK. In \autoref{fig:unit_no_sys}, we show the reduced HI densities, $\rho_{\rm HI} = 1+\delta_{\rm HI}$, of our Intensity map in a slice ($0 < z < 5$ ${\rm Mpc}/h$) of one of our four simulation boxes, \textsc{UNIT1}. Overlaid on this, we display the HI galaxy positions (with only 0.5 percent selected randomly shown for clarity) as yellow stars whose sizes are proportional to their HI masses. In \autoref{fig:unit_sys}, we show the same slice after applying observational systematics, described in the next section.

\subsection{Summary statistics}
\label{sec:summary_stats_unitsims}

In this work we consider the probability density function (PDF) smoothed at different radii $R$, $\mathcal{P}(R, \delta_{\rm HI})$, and the HI power spectrum, $P(k)$, as our set of summary statistics. 

To construct smoothed HI fields we use the \textsc{Pylians} smoothing library \citep{Pylians}, applying spherical top-hat filters of radius $R$. We then compute the corresponding smoothed PDFs, $\mathcal{P}(\delta_{\rm HI})$, using the \texttt{np.histogram} function from \textsc{numpy}, adopting 22 quantiles. Each quantile considers the same amount of overdensity voxels from our catalogues, that is, bins become wider in the tails such that we can get more stable results. 

We compute the (cross-)power spectra of our unsmoothed matter and HI density fields using \textsc{Pylians}, adopting linearly spaced wavenumber bins of width
$\Delta k = k_{\rm F}$,
between the fundamental mode
$k_{\rm F} = 2\pi/L_{\rm box}$,
and  $k_{\rm max} = 0.097\, h/{\rm Mpc}$. All HI intensity maps used in this work have $L_{\rm box} = 1000\,{\rm Mpc}/h$ and $L_{\rm cell} = 1\,{\rm Mpc}/h$.

\section{Observational systematics}
\label{sec: observational_systematics}

\subsection{Foreground removal}
\label{sec: foreground_removal}

Foregrounds are radio signals mainly coming from our galaxy due to synchrotron emission. Those signals are several orders of magnitude higher than the cosmological signal and share a similar frequency as the HI signal in our observer frame which leads to signal mixing. This is one of the main challenges of performing precision cosmology with Intensity Mapping. Methods to remove foregrounds rely on the fact that they are smooth in frequency. Then, the largest radial modes are progressively damped with an exponential suppression
\begin{equation}
\label{eqn:fkparallel}
    f\left(\frac{k_{\parallel}}{f_{\rm FG}}\right) = 1 - \exp\left(-\left[\frac{k_{\parallel}}{k_{\rm FG}}\right]^2 \right)\,,
\end{equation}
where $2\pi/k_{\rm FG}$ is the effective scale above which signal is suppressed. The largest cosmological modes will also be exponentially suppressed, since within this framework they cannot be disentangled with foregrounds. 
We parametrise $k_{\rm FG}= N_{\parallel} 2\pi/L_z$ with the radial length of the survey, $L_z$. \citet{Soares2021} found that $N_{\parallel} = 2$ emulates the FastICA foreground-cleaning method, developed in \cite{Hyvarinen1999}, removing $N_{\rm IC} = 4$ independent components. Considering the full SKA range $0.35 < z < 3$, we find $k_{\rm FG} = 3.64 \times 10^{-3}$ $h/{\rm Mpc}$.

This approximate prescription for the foreground removal neglects some effects like mode mixing induced by the instrument or residual contamination after cleaning. We do not attempt to model all effects explicitly but we instead build a controlled emulator that captures the dominant smooth-spectrum behaviour required for foreground removal.

The implementation of the foreground removal effect is described in the next subsection, where we describe the telescope beam effect and its implementation.

\subsection{Telescope beam}
\label{sec: telescope_beam}

The telescope beam effect is an observational systematic caused by the diffraction of incident light on the telescope antenna, resulting in a complex angular pattern of lobes. The strength of this effect depends on redshift $z$ and scales inversely with the antenna's diameter $D_{\rm dish}$. To first order, we approximate this effect as a Gaussian convolution with fixed width. The corresponding kernel in Fourier space is given by 
\begin{equation}
\label{eqn:bkperp}
    B (k_{\perp} R_{\rm beam}) = \mathcal{F}\left[\mathcal{N} (\mu,\sigma^2 = R_{\rm beam}^2)\right] (k_{\perp}R_{\rm beam}) = \exp\left[{-\frac{R_{\rm beam}^2 k_{\perp}^2}{2}}\right] \, ,
\end{equation}
where $R_{\rm beam}$ is the transverse smoothing scale, which is the comoving distance corresponding to an smoothing angle $\theta_{\rm FWHM} = \lambda_{\rm HI}(1+z)/D_{\rm dish}$, where $\lambda_{\rm HI} (1+z)$ is the redshifted HI line and $D_{\rm dish} = 15$m. 

For an SKA1-MID HI single-dish Intensity Mapping observation at $z_{\rm mid} = 1.321$ we obtain  $R_{\rm beam}= 38.45$ ${\rm Mpc}/h$ . 
We can apply this effect in our HI galaxy catalogues randomly displacing angular positions of HI galaxies according to a Gaussian distribution as in \citep[][]{Avila_BVG_2021}
\begin{equation}
\label{eqn:Gaussian}
    \mathcal{N} (\mu,\sigma^2 = R_{\rm beam}^2) = \frac{1}{\sqrt{2 \pi R_{\rm beam}^2}} \exp\left[{-\frac{(x-\mu)^2}{2 R_{\rm beam}^2}}\right] \,.
\end{equation}

This description accurately recovers the large-scale behavior, but it significantly overpredicts the power spectrum $P(k)$ for $k > 0.05 $ $h/{\rm Mpc}$ , due to the discreteness of our HI galaxy catalogue. To mitigate this overprediction one can increase the number density (e.g. for \( n_{50} = 50\, \bar{n}_{\rm HI} \) we achieve agreement up to \( k = 0.2\, h/\mathrm{Mpc} \), with errors smaller than 5 percent). Nevertheless, this method is increasingly computationally expensive and we still have some inaccuracies for the power spectrum at mildly non-linear scales.  We find that the most effective way to implement this systematic is in Fourier space, in analogy to how we model foreground subtraction in \autoref{sec: foreground_removal}. 

Below we show the process we follow when applying foreground removal and telescope beam together. We perform a discrete Fourier transform of the HI field, add the Fourier-space kernels with the distant observer approximation $k_{\parallel} = k_z$, $k_{\perp} = \sqrt{k_x^2+k_y^2}$ and then transform back. 
\begin{equation}
    \Delta T_{\rm HI,sys}(x,y,z) = \sum_{k_x,k_y,k_z} \Delta T_{\rm HI}(k_x,k_y,k_z) f(k_{\parallel}) B(k_{\perp}) e^{2\pi i\left(x k_x + y k_y + z k_z\right)/N } \, ,
\end{equation}
where $N = L_{\rm box}/L_{\rm cell} = 1000$, $k_{x,y,z} = 2\pi n/N$ $\left[h/\rm{Mpc}\right]$  with $n = [0, 1, \dots, \frac{N}{2}-1, -\frac{N}{2}, \dots, -1]$ and 
\begin{equation}
    \Delta T_{\rm HI}(k_x,k_y,k_z) = \sum_{x,y,z} \Delta T_{\rm HI}(x,y,z) e^{-2\pi i \left(x k_x + y k_y + z k_z\right) /N} \, .
\end{equation}

Returning now to \autoref{fig:unit_sys}, we see the effect of the combined observational systematics in our HI field, where we can see that small angular scale structures are suppressed, mainly due to the telescope beam effect. Nevertheless we still maintain great radial resolution. 

\subsection{Thermal noise}
\label{sec: thermal_noise}

\begin{figure}[t]
    \centering
    \begin{subfigure}[t]{0.48\textwidth}
        \centering
        \includegraphics[width=\textwidth]{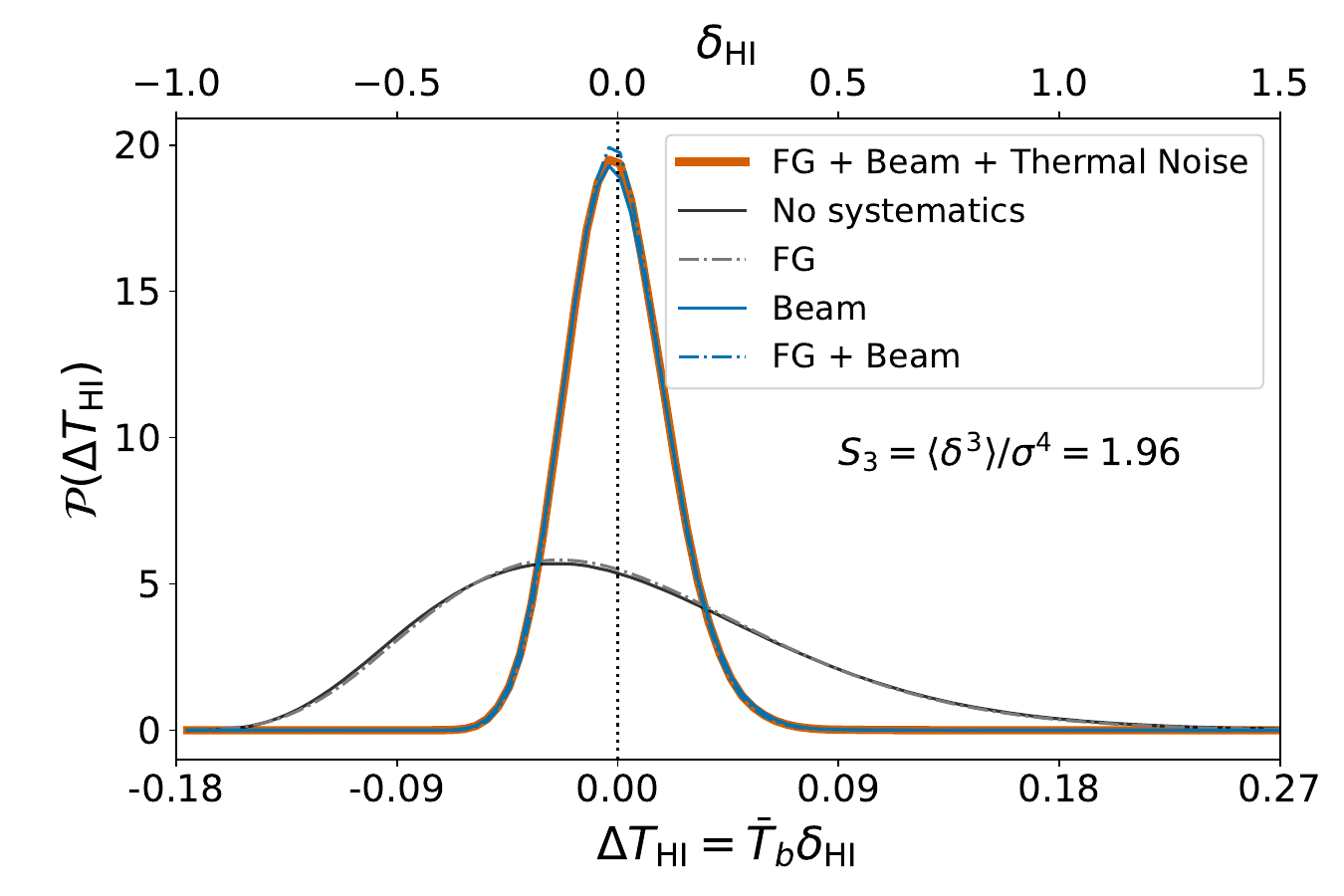}
        \caption{$R=15.5$ ${\rm Mpc}/h$}
        \label{fig:PDF_systematics_25}
    \end{subfigure}
    \hfill
    \begin{subfigure}[t]{0.48\textwidth}
        \centering
        \includegraphics[width=\textwidth]{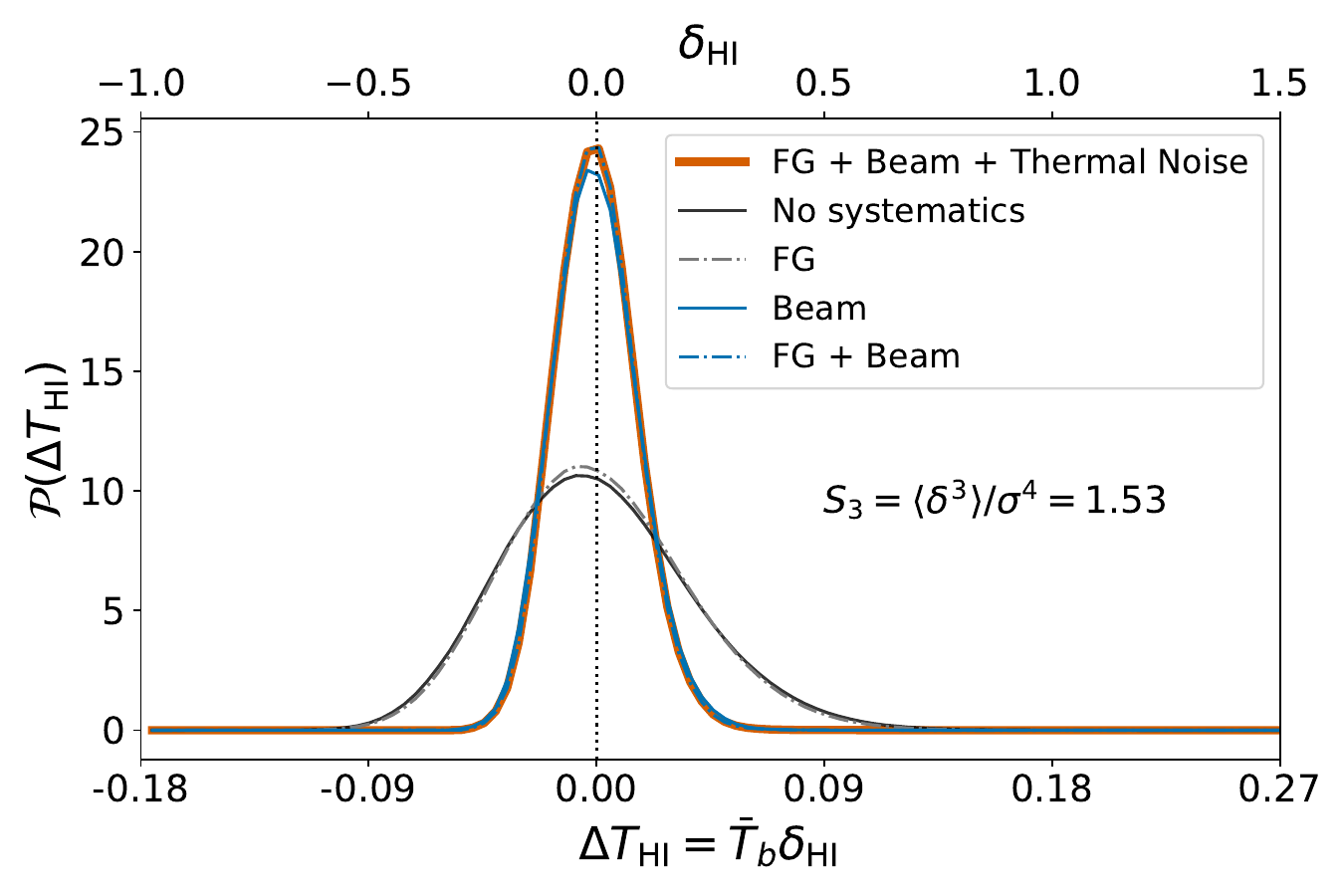}
        \caption{$R = 31.0$ ${\rm Mpc}/h$}
        \label{fig:PDF_systematics_50}
    \end{subfigure}
    \caption{Normalised one-point probability density function (PDF) of our HI Intensity Maps at redshift $z = 1.321$ smoothed with a top hat kernel with smoothing radius $R=15.5$ ${\rm Mpc}/h$ (left) and $R=31.0$ ${\rm Mpc}/h$ (right) as a function of the HI temperature $\Delta T_{\rm HI} = \bar{T}_b \delta_{\rm HI}$. Black solid lines represent the PDF without systematics, dot-dashed lines include the foreground removal effect and blue lines include the telescope beam effect. Finally, the red solid line include both systematics and thermal noise. The reduced skewness of the PDF with all systematics included is explicitly stated.}
    \label{fig:PDF_BVG}
\end{figure}

Thermal noise is an important source of noise for IM experiments. In our catalogues, we model it by adding a Gaussian random field with zero mean and variance $\sigma_{\rm pix}^2$ to the temperature fluctuation field.
\begin{equation}
\label{eqn:T_HI_noisy}
    \Delta T_{\rm HI-noisy,i} = \Delta T_{\rm HI,sys,i} + \mathcal{N}_i (\mu = 0, \sigma^2 = \sigma_{\rm pix}^2 ) \,,
\end{equation}
where $i$ runs over all cells of the HI catalogue. The pixel thermal noise $\sigma_{\rm pix}$ is computed from the system temperature as \citep{Battye2013}
\begin{equation}
\label{eqn:pixel_thermal_noise}
    \sigma_{\rm pix} = \frac{T_{\rm sys}}{\sqrt{\Delta f t_{\rm total} (\Omega_{\rm pix}/S_{\rm area}) N_{\rm dish} N_{\rm beam} }} = 0.45 \hspace{1mm} \rm{mK} \, ,
    \quad
    T_{\rm sys} = T_{\rm inst} + 60 \left(\frac{300 (1+z)}{\nu_0}\right)^{2.55} = 34.75 \hspace{1mm} \rm{K} \, .
\end{equation}

$T_{\rm inst} = 25\rm{K}$ is the instrument temperature for SKA1-MID and $\nu_0 = 1420\rm{MHz}$ is the rest-frame frequency of the neutral hydrogen hyperfine transition. The second term refers to the redshift-dependent sky temperature, which we evaluate at $z = 1.321$ as in all of our calculations.

The number of dishes of SKA1-MID is $N_{\rm dish}= 197$ and for a single dish observation the number of beams is $N_{\rm beams} = 1$. Usually, the pixel area is estimated as $\Omega_{\rm pix} \sim 1.13 \times \theta_{\rm FWHM}^2$ but our Intensity Maps are made of voxels with a section of $S_{\rm cell} = 1\,(${\rm Mpc}/h)$^2$, which corresponds to
\begin{equation}
\label{eqn:Omega_pix}
    \Omega_{\rm pix} = \left(\frac{1 {\rm Mpc}/h}{\frac{c}{H_0} \int_0^{z_{\rm mid}} \frac{dz}{E(z)}} \right)^2= 1.28 \times 10^{-7} \hspace{1mm} \rm{steradians} \,,
\end{equation}
where the denominator represents the comoving distance squared up to redshift $z_{\rm mid} = 1.321$. To compute the frequency shift, we determine the redshift bin width $\Delta z$ from the pixel radial length  
\begin{equation}
\label{eqn:r_pix}
    r_{\rm pix} = \int_{z_{\rm min}}^{z_{\rm max}} dz \frac{c}{H(z)} = 1 \hspace{1mm} \rm{Mpc}/h \, ,
\end{equation}
where $z_{\rm min} = z_{\rm mid} - \Delta z/2$ and $z_{\rm max} = z_{\rm mid}+ \Delta z/2$ are the limits of a redshift bin $\Delta z = 6.8 \times 10^{-4}$.

Thus, the frequency shift for this redshift slice can be computed as 
\begin{equation}
\label{eqn:Delta_f}
    \Delta f =  \frac{\nu_0}{\left(1+z_{\rm mid}\right)^2} \Delta z = 25 \hspace{1mm} \rm{MHz} \, .  
\end{equation}

We have that the channel width for SKA1-MID Band 1 ($0.35 < z < 3$) is 13.4 kHz, but in this work we use one- and two-point statistics evaluated at much larger radial (and angular) scales. We leave the design of angular one-point statistics leveraging the excellent radial resolution of the SKA IM survey for future work.

Finally, we consider a total observing time of $t_{\rm total} = 10000$  hours and observation area of $S_{\rm area} = 4 \pi f_{\rm sky}$, with an sky fraction $f_{\rm sky} = 0.3$.

\autoref{fig:PDF_BVG} shows the one-point PDF of the HI Intensity Maps considered in this work with and without the systematics introduced above, smoothed with a top hat kernel with radius of $R = 15.5$ ${\rm Mpc}/h$ (left) and $R = 31.0$ ${\rm Mpc}/h$ (right). Black lines represent the PDF without any kind of systematics, dashed lines include the foreground removal effect and blue lines include the telescope beam effect. Finally, HI PDFs including thermal noise on top of beam and foreground systematics are represented with red solid lines. We see that the telescope beam is the systematic that most strongly affects our statistic, noticeably diminishing the PDF variance due to the additional angular smoothing. In contrast, foreground removal slightly modify the PDF shape. Finally, although we have an important contribution of thermal noise in our original Intensity Maps with $L_{\rm cell} = 1$ ${\rm Mpc}/h$, its contribution is greatly reduced after spherical smoothing to .  As expected, the variance ratio between PDFs with and without beam contribution diminishes when we increase the smoothing radius. For spherical smoothing radii $R \gg R_{\rm beam}$ we expect a minimal contribution of the telescope beam, but those smoothing scales would be so large that they erase almost all non-Gaussian information. We note that PDF shapes contain non-Gaussian information that can still be exploited. The reduced skewness the PDF including all systematics is $S_3 = \langle \delta^3 \rangle/\sigma^4 = 1.96 (1.53)$ for $R = 15.5(31.0)$ $h/{\rm Mpc}$. From this point onward, we will consider PDFs computed from smoothed HI intensity maps, using three smoothing radii of $R = 15.5, 24.8,$ and $31.0$ ${\rm Mpc}/h$.

In this work, we model and test our summary statistics in real space, neglecting the effect of HI peculiar velocities. While redshift-space distortions (RSD) are an important ingredient for realistic cosmological analyses, we assess the potential improvement by non-Gaussian statistics in real space to study the impact of nonlinear evolution and observational systematics in a controlled setting. A more complete treatment including RSD is left for future work. For the PDF model, RSD effects could be absorbed into the dark matter variance, HI bias and shot noise contributions explained in the next section \citep[see e.g.][]{Uhlemann_2018,Leicht2019,ReppSzapudi2020}.

\section{Modeling and testing summary statistics}
\label{sec:model_test_summary_statistics}

In this section, we develop theoretical models for the HI Probability Density Function (PDF) and the HI Power spectrum, including the impact of telescope beam and foreground-removal systematics. We validate these models by comparing them to the simulation-based summary statistics (see \autoref{sec:summary_stats_unitsims}).

\subsection{HI probability density function (PDF)}
\label{sec:PDF}
In analogy to the approach used for modelling galaxy bias and shot noise in probability density functions \citep{Friedrich2022tracerPDF,Gould2025}, we model the HI PDF, $\mathcal{P}(\delta_{\rm HI})$, using the publicly available \href{https://github.com/OliverFHD/CosMomentum}{\textsc{CosMomentum}} code \citep{Friedrich2025jointPDF}
\begin{equation}
\label{eqn:HI_PDF_model}
\mathcal{P}(\delta_{\rm HI})
= \int \mathcal{P}(\delta_{\rm HI},\delta_m)\, d\delta_m
= \int \mathcal{P}(\delta_{\rm HI}|\delta_m)\, \mathcal{P}(\delta_m)\, d\delta_m \, ,
\end{equation}
where $\mathcal{P}(\delta_m)$ is the matter density PDF predicted by Large Deviation Theory (LDT) and $\mathcal{P}(\delta_{\rm HI}|\delta_m)$ is the conditional distribution of HI given the dark matter (DM) density contrast. To relate the HI overdensity $\delta_{\rm HI}$ to an HI temperature fluctuation $\Delta T_{\rm HI}$, we map the PDFs using conservation of probability such that $\mathcal P(\Delta T_{\rm HI} )= \mathcal P(\delta_{\rm HI}= \Delta T_{\rm HI}/\bar T_b )/\bar T_b$.

\subsubsection{Matter PDF model}
\label{sec:PDF_theory}

We briefly review the key ingredients for modelling of the probability density function of matter densities in spherical cells. The Large Deviation Principle is satisfied for a PDF, $\mathcal P(\delta)$, if it exists a rate function $\psi (\delta)$ such that $\lim_{\epsilon \to 0} \epsilon \ln \mathcal P(\delta) = - \psi (\delta) $ \citep{Bernardeau2016}. For Gaussian initial conditions, the PDF of the linear matter density contrast  in a sphere of radius $r$ is given in terms of a rate function that is quadratic in the linear density 
\begin{equation}
    \mathcal P(\delta_{\rm L}) =  \sqrt{\frac{1}{2\pi \sigma_{\rm L}^2 (r)}} \exp \left[-\frac{\delta_{\rm L}^2}{2\sigma_{\rm L}^2(r)}\right], \quad \Psi_{\rm L}(\delta_{\rm L}) = \frac{\psi_{\rm L}(\delta_{\rm L})}{\sigma_{\rm L}^2(r)} = \frac{\delta_{\rm L}^2}{2\sigma_{\rm L}^2(r)}\,,
\end{equation}
where the linear variance for a particular smoothing radius, $\sigma_{\rm L}^2 (r)$, is given in terms of the linear power spectrum, $P_{\rm L} (k)$, weighted by the spherical top-hat window function in Fourier space
\begin{equation}
    \sigma_{\rm L}^2 (r) = \int \frac{dk}{2\pi^2} P_{\rm L} (k) k^2 W_{\rm TH}^2 (kr),\quad  W_{\rm TH} (kr) =3 \frac{\sin \left(kr\right) - kr \cos \left(kr\right)}{\left(kr\right)^3}\,.
\end{equation}

The next step is to construct a PDF for the non-linear matter density $\delta_m$. The rate function of the late-time matter density contrast, $\psi (\delta_m)$, can be constructed as a function of $\psi_{\rm L} (\delta_L)$ using the contraction principle of LDT
\begin{equation}
    \psi (\delta_m) = \inf_{\delta_{\rm L}: \Phi(\delta_{\rm L}) = \delta_m}  \psi_{\rm L} (\delta_{\rm L}) = \psi_{\rm N} (\delta_{\rm L}^{\rm SC}(\delta_m)) \, ,
\end{equation}
where $\Phi$ is a continuous mapping that can be approximated by the spherical collapse (SC) $\Phi = \Phi_{\rm SC}$, since this provides the most probable evolution between the linear and non-linear densities in spherical regions \citep{Valageas2002}. The spherical configuration minimizes the potential energy required to reach the target density contrast, making it the most likely path in the gravitational dynamics. The right-hand side term takes into account that $\Phi_{\rm SC}$ is a monotonic function between $\delta_L$ and $\delta_m$. The new rate function can be expressed as
\begin{equation}
    \label{eqn:rate_function_psi}
    \Psi(\delta_m) = \frac{\sigma_{\rm L}^2(z,R)}{\sigma_{\rm NL}^2(z,R)} \frac{\delta_{\rm L}^{\rm SC}(\delta_m)^2}{2 \sigma_{\rm L}^2 (z,r)}\,,
\end{equation}
where we relate the initial radius, $r$, and the final radius, $R$, as $r=R(1+\delta_m)^{1/3}$ taking into account mass conservation. We included the non-linear variance $\sigma_{\rm NL}^2$, which is calculated by integrating the non-linear power spectrum obtained using \textsc{halofit} fitting function \citep{Takahashi_2012}. If instead we use the  non-linear power spectrum given by \textsc{hmcode-2020} \citep{Mead_2021}, we find very small differences that slightly intensify when decreasing $R$. We see up to 1 percent discrepancy for the modelled PDFs with the smallest smoothing scale considered in this work.

We can obtain the cumulant generating function as the Legendre transform of the rate function, 
\begin{equation}
    \phi(\lambda) = \log E[e^{\lambda \delta_m}] = \log \int d \delta_m \mathcal P(\delta_m) e^{\lambda \delta_m} \sim \log \int d \delta_m  e^{(\lambda  \delta_m-\psi/\sigma_{\rm NL}^2)} = \frac{\rm{sup}_{\delta_m} (\lambda \sigma_{\rm NL}^2 \delta_m - \psi (\delta_m))}{\sigma_{\rm NL}^2} \, ,
\end{equation}
where we rely on the saddle-point approximation for small values of the variance $\sigma_{\rm NL}^2 (R) \ll 1$ \citep{Uhlemann_2016}. That is, the integral is dominated by the value of $\delta_m$ that maximizes the exponent, so the result of the integral is approximately given by the supremum of the exponent. This approximation holds for the smoothing scales considered in this work. Finally, the probability density function of matter densities can be obtained taking the inverse Laplace transform of the cumulant generating function,

\begin{equation}
    \mathcal P(\delta_m) = \int_{-\infty}^{\infty} \frac{d\lambda'}{2\pi} \exp\left[ \phi (i\lambda')-i \lambda' \delta_m\right] \, ,
\end{equation}
where $i \lambda' = \lambda$. In Appendix \ref{ap:ap2}, we show that the matter PDF predicted by our model agrees extremely well with that measured from the UNIT dark matter particles, well within the $1\sigma$ error bars, for three smoothing scales ($R = 15.5$, $24.8$, and $31.0$  $\rm{Mpc}/h$). We also validate the PDF prediction including observational systematics against simulations.

\subsubsection{HI-Dark matter conditional distributions}
\label{sec:joint_PDF}

\begin{figure}[b]
    \centering
    \begin{subfigure}[t]{0.48\textwidth}
        \centering
        \includegraphics[width=\textwidth]{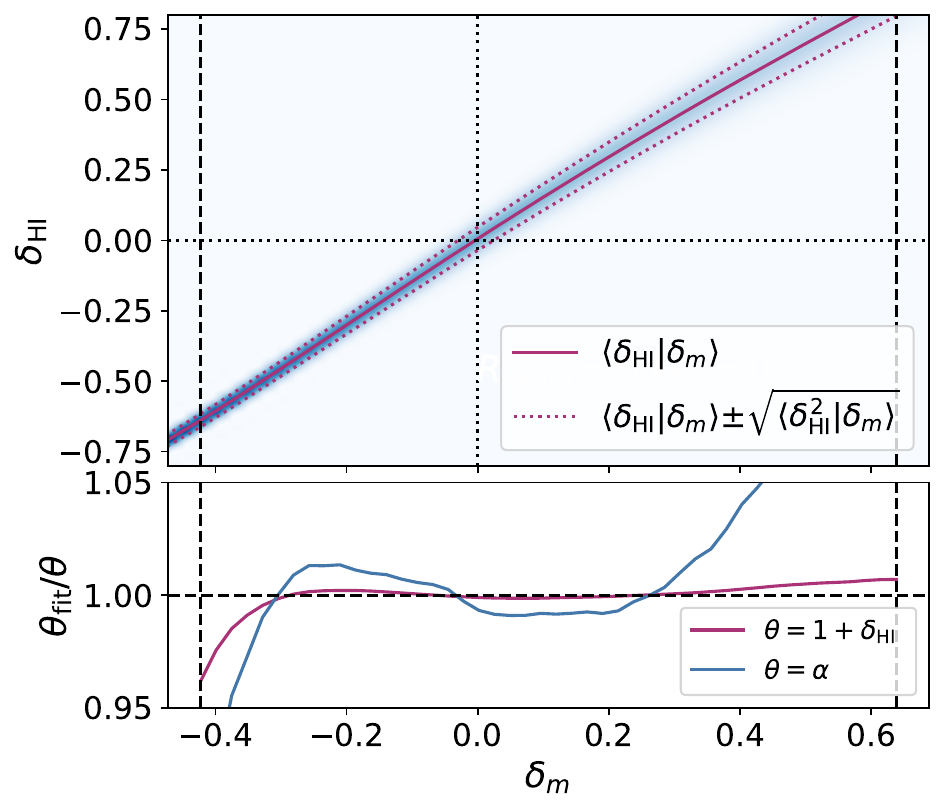}
        \caption{No observational systematics}
        \label{fig:HI_DM_no_sys}
    \end{subfigure}
    \hfill
    \begin{subfigure}[t]{0.48\textwidth}
        \centering
        \includegraphics[width=\textwidth]{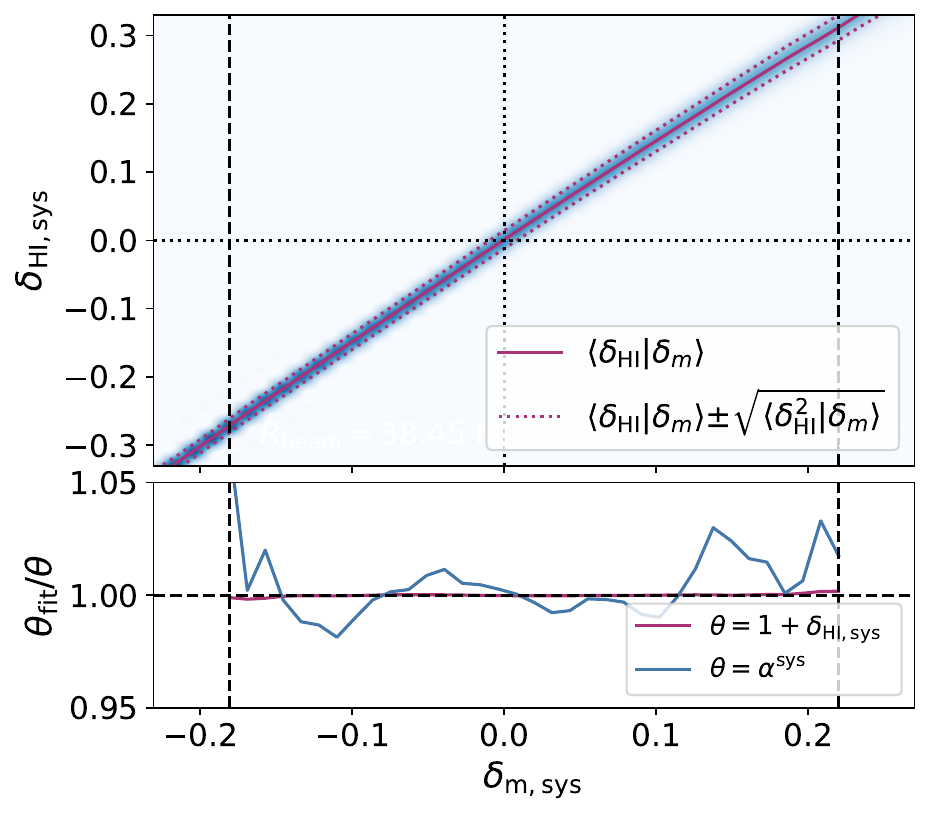}
        \caption{Considering observational systematics}
        \label{fig:HI_DM_sys}
    \end{subfigure}
    \caption{\textbf{Top: } Conditional distribution of HI given dark matter density contrast, computed at a smoothing scale of $R = 24.8$ ${\rm Mpc}/h$. The mean HI overdensity and the scatter, $ \langle \delta_{\rm HI} | \delta_m \rangle \pm \sqrt{\langle \delta_{\rm HI}^2 | \delta_m \rangle}$, are represented as magenta solid and dashed lines, respectively.
    \textbf{Bottom: } Ratio between the best-fit quadratic models for the mean HI density and the same quantity obtained from simulations (magenta) as well as the ratio of the shot noise (blue), all fitted in the $0.1-99.9$ percent range of the matter PDF (see black dashed vertical lines). Bias and shot noise best fit values are listed in Table \ref{tab:bias_and_shotnoise_fits}. }
    \label{fig:2D_distribution_HI_DM}
\end{figure}


To model the conditional PDF, $\mathcal P(\delta_{\rm HI}|\delta_m)$, we first examine the relation between neutral hydrogen and dark matter using the matter and HI catalogues from the UNIT simulations. We focus on the conditional one-point distribution of HI given matter density for different smoothing radii, $R$.

First, we will  show the relation before including  telescope beam and foreground-removal. We can extract from $\mathcal P(\delta_{\rm HI}|\delta_m)$ its first two moments, the conditional mean $\langle\delta_{\rm HI}|\delta_m\rangle$ and variance $\langle\delta_{\rm HI}^2|\delta_m\rangle_c$ for different smoothing scales, which set the HI-matter connection for the HI PDF. From those, we construct a non-Poissonian conditional distribution through a rescaling of the well-known Poisson distribution  
\begin{equation}
\mathcal P(N_{\rm HI}|\delta_m) = \mathcal P_{\rm P}\left(\tilde N_{\rm HI}=\!\frac{N_{\rm HI}}{\alpha(\delta_m)};\overline{\tilde N}_{\rm HI}=\frac{\overline{N}_{\rm HI}(\delta_m)}{\alpha(\delta_m)}\right) \frac{1}{\alpha(\delta_m)}\, ,   
\end{equation}
where $\bar{N}_{\rm HI} =\bar{n}_{\rm HI,eff} \times V(R)$ is the effective mean number of HI tracers (see \autoref{eqn:n_hi_eff}) enclosed in a spherical volume $V$ with radius $R = 15.5, 24.8, 31.0$ $\rm{Mpc}/h$. When we include systematics then we replace $V$ by the effective volume of the cells, given by
\begin{equation}
V_{\rm eff}(R,R_{\rm beam},k_{\rm FG})=\left[\int \frac{d^3k}{(2\pi)^3}
\,W_{\rm TH}^2(kR)\,
B^2(k_\perp R_{\rm beam})f^2 \left(\frac{k_\parallel}{k_{\rm FG}}\right)\,
\right]^{-1} ,
\end{equation}
which increases it by a factor 32 to 9 compared to the sphere volume from small to large smoothing scales.
For details including an explicit expression for the systematics including the beam we refer to \autoref{eq:Veff_beam} in \autoref{ap:ap1}. 

For the conditional mean, we adopt a second order Eulerian bias expansion as  considered in \cite{Friedrich2022tracerPDF,Gould2025,Friedrich2025jointPDF}
\begin{equation}
\label{eqn:delta_HI}
    \langle \delta_{\rm HI} | \delta_{\rm m} \rangle = b_1 \delta_m + \frac{b_2}{2} \left(\delta_m^2 - \sigma_m^2\right) \, .
\end{equation}

When modelling the bias between the HI and matter density with systematics we replace ($b_1$,$b_2$) by ($b_{1}^{\rm sys}$,$b_{2}^{\rm sys}$) for clarity. We measure the density-dependent stochasticity considering the conditional variance ratio and parametrise it with a second order expansion as adopted in \cite{Friedrich2022tracerPDF,Britt2024,Gould2025}
\begin{equation}
\label{eqn:alpha_HI}
    \alpha (\delta_m) = \frac{\bar{N}_{\rm HI}  \langle \delta_{\rm HI}^2 | \delta_{\rm m} \rangle_c}{ 1+\langle \delta_{\rm HI} | \delta_{\rm m} \rangle} =\alpha_0 + \alpha_1 \delta_m + \alpha_2 \delta_m^2 \, .
\end{equation}

When considering the stochasticity between the HI and matter density with systematics we replace ($\alpha_0$,$\alpha_1$,$\alpha_2$) by ($\alpha_0^{\rm sys}$,$\alpha_1^{\rm sys}$,$\alpha_2^{\rm sys}$) for clarity. In the case of a Poissonian distribution this ratio would be unity. We report the best-fit values for the quadratic bias and stochasticity parameters in Table~\ref{tab:bias_and_shotnoise_fits}. We show the explicit shape of $\alpha(\delta_m)$ in Appendix \ref{ap:HIstochasticity}. We note that the stochasticity slope increases when systematics are included ($\alpha_1^{\rm sys}$,$\alpha_2^{\rm sys} \gg \alpha_1,\alpha_2$ ) which is mainly because $\alpha_1^{\rm sys}, \alpha_2^{\rm sys}$ inherit the variance rescaling of the overdensity: $\delta_{\rm m,sys}/\sigma_{\rm m,sys} \sim \delta_{\rm m}/\sigma_{\rm m}$.  

Instead of using the parametrised quadratic Eulerian bias and shot noise, \textsc{CosMomentum} can also predict the tracer PDF if we provide directly the tabulated conditional mean and variance ratio as a function of matter density contrast, which serves as an intermediate step to check the accuracy of the HI PDF based on bias and shot noise parametrisations.

\begin{table}
  \centering
  \begin{tabular}{|c|c|c|c|c|c|c|c|c|c|}
      \hline
       ~ & $b_1^{P(k)}$ & $R \left[\rm{Mpc}/h\right]$ & $b_1$ & $b_2$ & $\alpha_0$ & $\alpha_1$ & $\alpha_2$ & $S_{\rm 3,m}^{\rm sim}$ & $S_{\rm 3,m}^{\rm th}$ \\ \hline \hline
      ~ & & 15.5 & $1.509$ & $-0.265$ & $1.160$ & $0.838$ & $2.829$ & $3.23$ & $3.20$  \\
      DM and HI without systematics & $1.471$ & 24.8 & $1.480$ & $-0.316$ & $1.762$ & $0.741$ & $5.304$ & $2.96$ & $2.95$ \\
      ~ & & 31.0 & $1.469$ & $-0.329$ & $2.347$ & $0.196$ & $7.452$ & $2.83$ & $2.82$ \\
      \hline
      \hline
      &  &  & $b_{1}^{\rm sys}$ & $b_{2}^{\rm sys}$ & $\alpha_0^{\rm sys}$ & $\alpha_1^{\rm sys}$ & $\alpha_2^{\rm sys}$ & & \\ \hline
      \hline
      ~ & & 15.5 & $1.474$ & $-0.459$ & $1.650$ & $2.913$ & $13.007$ & $4.34$ & $4.47$\\
      DM and HI with systematics & $1.467$ & 24.8 & $1.471$ & $-0.438$ & $ 1.873$ & $  2.850$ & $15.428$  & $3.82$ & $4.16$ \\
      ~ & & 31.0 & $1.470$ & $-0.428$ & $2.021$ & $2.638$ & $15.990$ & $3.58$ & $3.98$ \\
      \hline
    \end{tabular}
	\caption{Bias and shot noise best fit parameters for the three smoothing scales we consider in this work, linear bias $b_1$ fit to the power spectrum $P(k)$ and reduced skewness, $S_{\rm 3,m} = \langle\delta_{\rm m}^3\rangle/\sigma_{\rm m}^4$ for dark matter. The offset in the skewness values for the simplified PDF systematics modelling is explained in Appendix~\ref{ap:ap1}.}
	\label{tab:bias_and_shotnoise_fits}
\end{table}

In the upper panel of \autoref{fig:HI_DM_no_sys} we show the conditional PDF, $\mathcal{P}(\delta_{\rm HI} | \delta_m)$ for spheres of radius $R = 24.8$ ${\rm Mpc}/h$ without including observational systematics. The magenta solid line represents the conditional mean while magenta dotted lines represent the scatter, proportional to the shot noise. In the bottom panel we show the accuracy of the fits for the conditional mean (magenta line) and shot noise (blue line) applying \autoref{eqn:delta_HI} and \autoref{eqn:alpha_HI}, respectively. Although the shot noise residuals are larger, errors in the conditional mean (bias) have a much greater impact on the computation of the HI PDF. Black dashed vertical lines indicate our fitting range, where we take into account matter density bins between 0.1-99.9 percent of the matter cumulative density function (CDF). As expected, we find an approximately linear relationship, but also a preference for lower HI density in regions where the dark matter overdensity is far from zero ($|\delta_m| \gg 0$), which yields a negative value of $b_2$ (see Table~\ref{tab:bias_and_shotnoise_fits}).

Best-fit parameters are shown in Table~\ref{tab:bias_and_shotnoise_fits}. Our conditional variance fit further shows that the HI distribution is slightly super-Poissonian and depends strongly on the smoothing scale. As anticipated, the linear bias measured from the conditional mean at increasing smoothing scale approaches the linear bias inferred from the power spectrum. We find qualitatively similar results for our HI tracers and halos at similar redshifts \citep{Gould2025}, the former being more biased and confirm the necessity of nonlinear bias and  stochasticity models with significant scale-dependence.

Having explained how to obtain the matter PDF and the conditional PDF, they can now be combined to obtain the HI PDF using \autoref{eqn:HI_PDF_model}. To model the HI PDF including telescope beam and foreground-removal effects, we  need to modify both ingredients. To apply systematics in the dark matter PDF, we use  linear and nonlinear power spectra with systematics (see \autoref{eqn:Seff_k}) to compute the corresponding linear and nonlinear variances appearing in the rate function \autoref{eqn:rate_function_psi}. In  particular, this rescales the non-linear variance of the matter PDF (see Appendix \ref{sec: variance_computations} for a comparison of the matter and HI non-linear variances with and without systematics with simulations) and the reduced skewness (see Appendix \ref{sec:skewness} for a description of how observational systematics impact the third moment of the matter PDF). To build the conditional distribution, we compute the conditional mean and variance of the joint PDF that links DM and HI, incorporating the same observational systematics in both fields. 

\autoref{fig:HI_DM_sys} presents the conditional PDF with telescope beam and foreground-removal systematics included. The corresponding conditional mean and variance fits reveal several trends. The linear response parameter $b_{1}^{\rm sys}$  is slightly reduced compared to $b_1$, and the conditional variance ratio increases slightly. This reduction reflects the angular smoothing used to model the beam, which suppresses small-scale, non-correlated fluctuations between HI and dark matter. In the lower panel we show the corresponding ratios. Note that fitting limits represented as black dashed vertical lines are placed nearer to $\delta_{\rm m,sys} = 0$, since large fluctuations of matter and HI overdensities are reduced mainly by the beam smoothing when systematics are applied. 
  
\subsubsection{Validating the HI PDF model with UNIT simulations}
\label{sec:PDF_theory_and_simulations}

 \begin{figure}[t]
    \centering
    \begin{subfigure}[t]{0.49\textwidth}
        \centering
        \includegraphics[width=\textwidth]{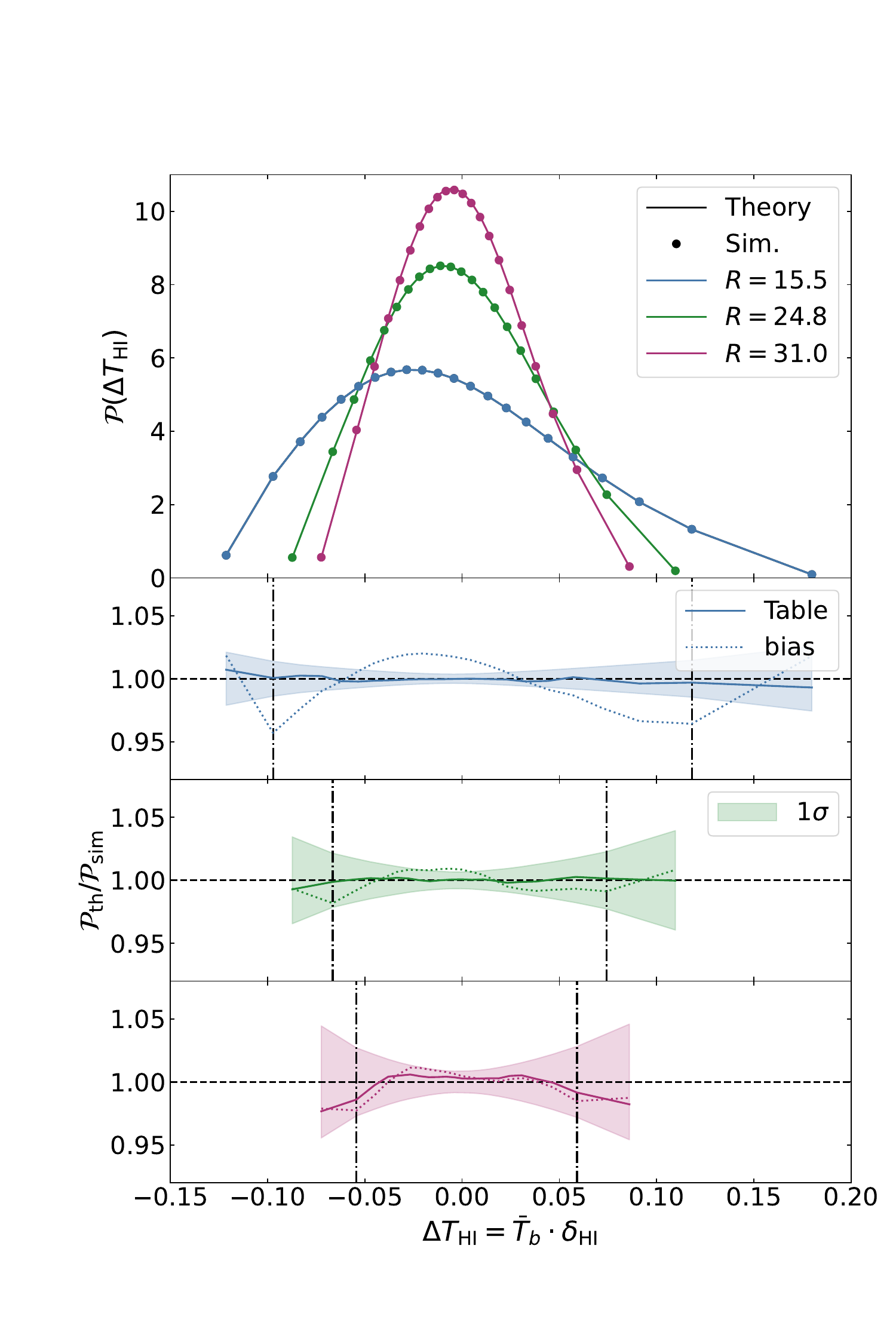}
        \caption{No observational systematics}
        \label{fig:PDF_CosMomentum}
    \end{subfigure}
    \hfill
    \begin{subfigure}[t]{0.49\textwidth}
        \centering
        \includegraphics[width=\textwidth]{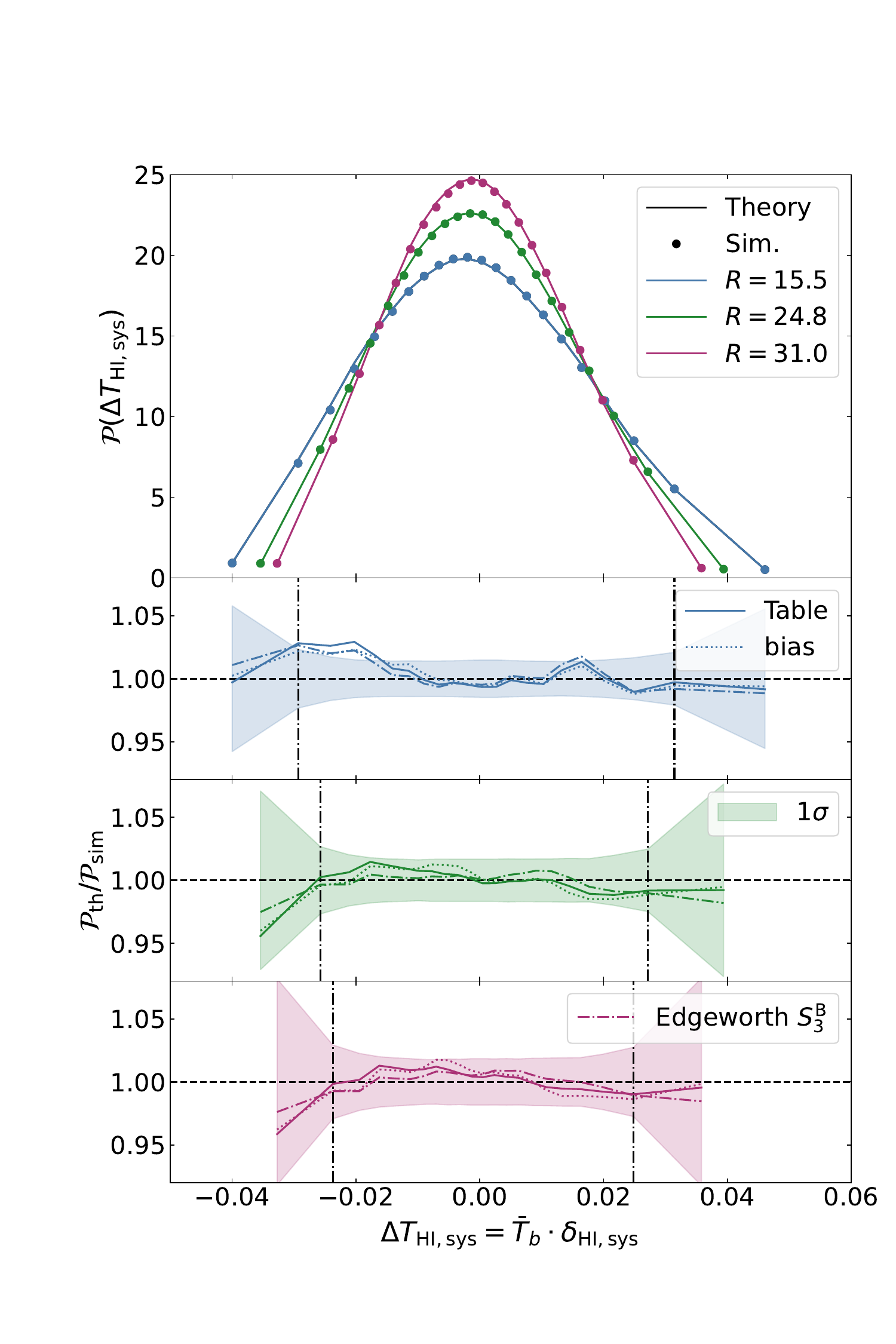}
        \caption{Considering observational systematics}
        \label{fig:PDF_CosMomentum_sys}
    \end{subfigure}
    \caption{\textbf{Top:} Simulation-based HI PDFs with smoothing radii of $R = 15.5$, $24.8$, and $31.0$ ${\rm Mpc}/h$ from HI Intensity maps at $z = 1.321$ are shown as points, plotted as a function of the temperature fluctuations $\Delta T_{\rm HI}$. The corresponding theory PDFs obtained with \textsc{CosMomentum}, are shown as solid lines.   \textbf{Bottom:} Ratio between the theoretical and simulated PDFs, with the jackknife-based $1\sigma$ uncertainty shown as shaded regions. On the right, the systematics skewness-corrected predictions are shown as dot-dashed lines. Dotted lines use the quadratic best fits for the conditional mean and variance instead of using directly those from simulations, represented as solid lines. Vertical black dashed lines indicate the scale cuts applied for each smoothing scale $R$ which exclude the first and last quantiles.}
    \label{fig:PDF_CosMomentum_no_sys_and_sys}
\end{figure}
In this subsection, we assess the performance of our theoretical model by comparing its predictions with the HI PDF measured in simulations.

In \autoref{fig:PDF_CosMomentum} we compare model and simulated HI PDFs as a function of the temperature fluctuations $\Delta T_{\rm HI}$, for three smoothing scales: $R=15.5, 24.8, 31.0$ ${\rm Mpc}/h$ without including observational systematics. At the top we show the simulated and model PDFs binned in quantiles as data points and solid lines, respectively. Model PDFs are originally produced with very fine binning and are rebinned to quantiles by computing the mean value of the probability density in each interval. At the bottom we show the corresponding ratios for each smoothing scale, where shaded regions represent Jackknife 1-$\sigma$ errors (see section \ref{sec:covariance} for the covariance computation details). Modelled PDFs using directly the simulated conditional mean and variance to provide the matter-HI connection are represented as solid lines, which show an excellent agreement between theory and simulations. Dotted lines show modelled PDFs where we use instead a second order Eulerian bias (see \autoref{eqn:delta_HI}) and second order expansion for the shot noise (see \autoref{eqn:alpha_HI}). We find up to 4 percent deviations for the smallest scale,  $R=15.5$ ${\rm Mpc}/h$, mostly arising from the propagation of errors recovering the conditional mean (see the lower panel of \autoref{fig:HI_DM_no_sys}). This implies that higher order bias expansion may be needed to better fit the conditional moments\footnote{Second order Lagrangian bias and Gaussian Lagrangian bias expansions \citep{Gould2025} yielded slightly better results but still outside error bars}. As expected, the agreement between theory predictions and simulations increases with the smoothing scale $R$, since nonlinearities decrease.

In \autoref{fig:PDF_CosMomentum_sys} we show again the simulated and the theoretical PDFs, in this case including systematics, when using the conditional mean and variance directly from simulations (solid lines) and  the bias and shot noise fits (dotted lines). Overall the variance decreased mainly due to the telescope beam effect. Solid and dotted lines show better agreement than in \autoref{fig:PDF_CosMomentum}, since the smoothing introduced by systematics suppresses higher-order bias terms.  All scales show a skewness discrepancy as our model assumes a specific scaling of skewness with variance, while anisotropic systematics introduce deviations. For the smallest scale we have errors up to 3 percent. This can be traced back to discrepancies in the dark matter reduced skewness values when considering systematics through the power spectrum modification (see the last two columns in  \autoref{tab:bias_and_shotnoise_fits}), which we explain in Appendix \ref{ap:ap1}. While their visual impact on the dark matter PDF shape around the peak is subtle (see Figure~\ref{fig:DM_PDF_CosMomentum_no_sys_and_sys}), the convolution required to obtain the HI PDF enhances this effect. We also provide an Edgeworth correction for the skewness of the dark matter PDF with systematics to match the tree-level skewness (see equation~\eqref{eqn:DM_PDF_sys_Edgeworth_2} in Appendix \ref{sec:skewness} for more details). Convolving this corrected PDF with the conditional distribution yields our new model for the HI PDF, shown as dashed-dotted lines. This tends to improve all smoothing scales, while differences for the small scales where the tree-level approximation is not sufficient are expected.

In summary, we find that our modelled PDFs agree with simulations but we need to better understand the skewness term in the presence of systematics specially for small smoothing scales. 

\subsection{HI Power spectrum}
\label{sec:power_spectrum_theory}
In this section, we present a theoretical model for the HI power spectrum that incorporates the effects of the telescope beam, foreground removal, shot noise, and thermal noise.

\subsubsection{Modelling telescope beam and foreground-removal systematics}

We use \textsc{hmcode-2020} \citep{Mead_2021} to predict the matter power spectrum, $P_{\rm m}(k)$, which is implemented in \textsc{camb}. We consider a linear bias model to describe the HI power spectrum at mildly non-linear scales. In particular, we fit $b_1$ using the scales $0.01 < k $ $[\text{Mpc}/h] < 0.1$  of the bias function, which is obtained dividing the reference UNIT HI-matter cross power spectrum and matter power spectrum (see \autoref{fig:bias_comparison}). 

The theoretical power spectrum on mildly nonlinear scales $k < k_{\rm max}$ is modelled as
\begin{align}
\label{eqn:pk_theory_and_sys}
P_{\rm theory}(k) &= b_1^2\,P_{\rm m}(k) + P_{\rm shot} \, ,
\end{align}
where the thermal noise contribution decides at which $k_{\rm max}$ we should cut, as we explain in the next subsection. The combined effect of both telescope beam and foreground-removal systematics is added by considering the following integral of the telescope beam and foreground removal kernels squared
\begin{equation}
\label{eqn:Seff_k}
    S_{\rm eff} (k; R_{\rm beam}, k_{\rm FG}) = \frac{1}{2} \int_{-1}^1 B^2 \left(k\sqrt{1-\mu^2} R_{\rm beam}\right) f^2 \left(\frac{k\mu}{k_{\rm FG}}\right) d\mu \, ,\quad P_{\rm theory,sys}(k) = 
P_{\rm theory}(k)S_{\rm eff}(k)\, ,
\end{equation}
where we used $k_\perp = k\sqrt{1-\mu^2}$ and $k_{\parallel}=k\mu$, with $\mu = \cos \theta$ as the cosine of the angle between the 
wavevector and the line of sight. We do not describe small scales in this power spectrum model since the thermal noise becomes dominant there. Also shot noise becomes important at mildly non-linear scales. Thermal and shot noise power spectra are estimated in the next subsection.

\subsubsection{Modelling shot noise and thermal noise}
\label{sec:shotnoise_and_thermalnoise}
We approximate the HI shot noise power spectrum as \citep{Wolz_2018}
\begin{equation}
\label{eqn: shotnoise_pk_constant}
    P_{\rm shot} = \frac{\int d\log_{10}(m) n(\log_{10}m) \langle M_{\rm HI} (\log_{10}m) \rangle ^2}{\left[\int d\log_{10}(m) n(\log_{10}m) \langle M_{\rm HI} (log_{10}m) \rangle \right]^2} = \frac{\langle M_{\rm HI}^2\rangle}{\bar{n}_{\rm HI} \langle M_{\rm HI}\rangle^2} = 77.4 \hspace{1mm} \left[\rm{Mpc}/h\right]^3 \, ,
\end{equation}
where $m$ is the halo mass, $n(\log_{10}m)$ is the halo mass function and $\langle M_{\rm HI} (\log_{10}m) \rangle$ is the HI mass distribution as a function of the halo mass. This prescription for the shot noise assumes a modified Poissonian sampling of HI in the dark matter field in the form of white noise. We can define an effective number density, $\bar{n}_{\rm HI,eff}$, that takes into account HI mass weighting for each galaxy

\begin{equation}
\label{eqn:n_hi_eff}
    \bar{n}_{\rm HI,eff} = \bar{n}_{\rm HI} \frac{\langle M_{\rm HI}\rangle^2}{\langle M_{\rm HI}^2\rangle} \, .
\end{equation}

The shot noise only affects the power spectrum at non-linear scales that may not be accessible due to the telescope beam smoothing. Nevertheless, we will include it in our theoretical estimation of the power spectrum. 

Now we should also take attention to the thermal noise of the instrument, which is the dominant noise contribution for those type of observations. Its power spectrum, $P_{\rm thermal} (k)$, increases notably when we push to larger wavevectors, which will force us to impose a maximum wavenumber, $k_{\rm max}$, in our analysis.

The thermal noise power spectrum is given in terms of the pixel noise, $\sigma_{\rm pix}$, the pixel volume, $V_{\rm pix}$, \citep{Battye2013} and the telescope beam kernel
\begin{equation}
\label{eqn: thermal_noise_pk}
    P_{\rm thermal} (k) = \sigma_{\rm pix}^2 V_{\rm pix} B(k)^{-2} \, ,
\quad
    V_{\rm pix} = \Omega_{\rm pix} \int_{z_{\rm min}}^{z_{\rm max}} dz \frac{c r(z)^2}{H(z)^2}\,, 
\end{equation}
where we approximated the original $k_\perp$-dependence of the telescope beam kernel $B(k_\perp)$ to a $k$-dependence, $B(k)$, since the expected radial resolution of IM surveys (and specially in SKA) is exquisite. We calculate $P_{\rm thermal} (k)$ considering $f_{\rm sky} = 0.3$ and $t_{\rm total} = 10000$ hours. We selected a scale cut for our signal HI power spectrum $P_{\rm HI, sys} (k)$ to stay significantly above the thermal noise contribution
\begin{equation}
    k_{\rm max} = \max \{ k \in \mathbb{R}^+ | 3 P_{\rm thermal}(k) < P_{\rm HI, sys}(k) \} = 0.097 \hspace{1mm} h/{\rm Mpc} \,.
\end{equation}

$k_{\rm max}$ mildly depends on $f_{\rm sky}$ and $t_{\rm total}$. In particular, for $f_{\rm sky} = 0.7$ and $t_{\rm total} = 5000$ hours we would obtain $k_{\rm max} = 0.091$ $h/{\rm Mpc}$, which would remove one bin in our analysis. 

\subsubsection{Testing model HI power spectrum with UNIT simulations}
\label{sec:pk_theory_and_simulations}
\begin{figure}[htbp]
    \centering
    \begin{subfigure}[t]{0.48\textwidth}
        \centering
        \includegraphics[width=\textwidth]{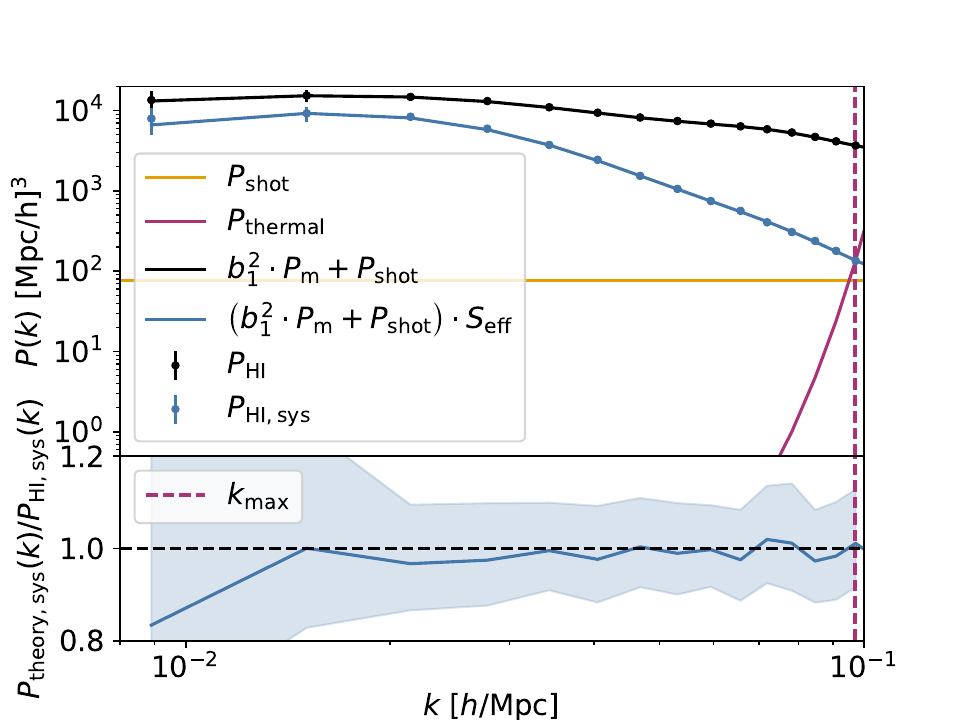}
        \caption{}
        \label{fig:pk_plot}
    \end{subfigure}
    \hfill
    \begin{subfigure}[t]{0.48\textwidth}
        \centering
        \includegraphics[width=\textwidth]{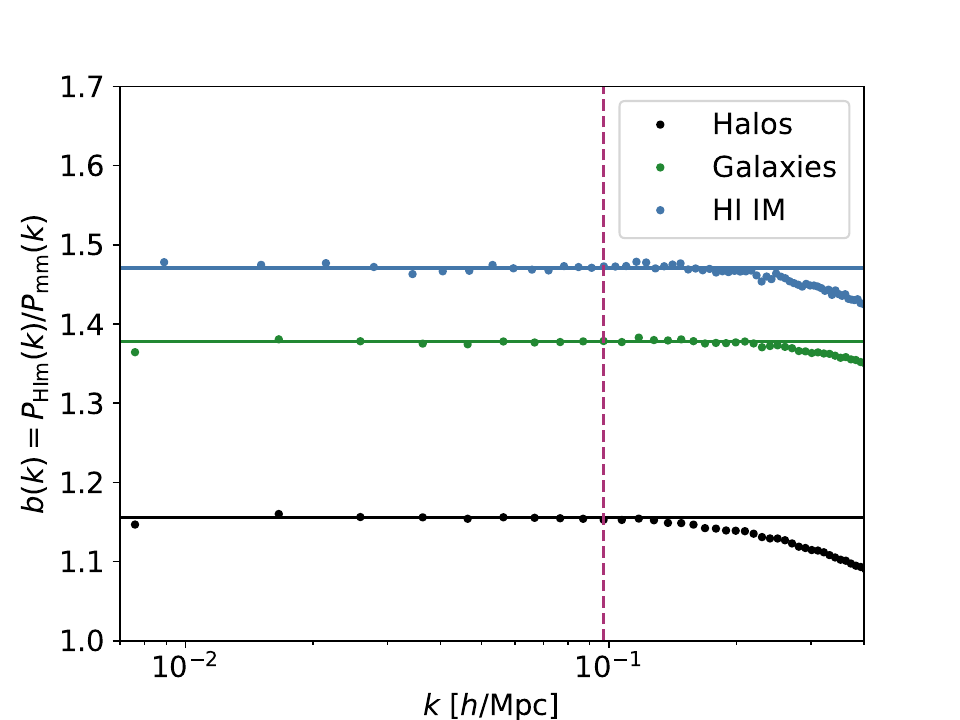}
        \caption{}
        \label{fig:bias_comparison}
    \end{subfigure}
    \caption{\textbf{Top left:} Comparison between UNIT and model HI power spectra. The shot noise constant power spectrum, computed using equation \ref{eqn: shotnoise_pk_constant}, is shown as an orange horizontal solid line. The thermal power spectrum, obtained from equation \ref{eqn: thermal_noise_pk}, is shown as a magenta line. UNIT HI power spectra without and with observational systematics are represented as black and blue points with $1-\sigma$ errorbars, respectively. Error bars are extracted using Jackknife resampling covariance (see equation \ref{eqn:Jack_c}). Finally, the theoretical estimates of the power spectrum are represented as solid lines. \textbf{Bottom left:} Ratio between the theoretical and simulated power spectra. Black and blue shaded regions represent the $1-\sigma$ errorbars without and with systematics, respectively. The magenta dashed vertical line represents the maximum scale we consider in our power spectrum analysis. \textbf{Right: } Bias functions for the haloes, HI galaxies and the HI Intensity Map in black, green and blue points, respectively. best fits for the linear bias are shown as accordingly colored horizontal solid lines.}
    \label{fig:power_and_bias}
\end{figure}

In \autoref{fig:pk_plot}, we test the accuracy of the theoretical model for the power spectrum presented in Section \ref{sec:power_spectrum_theory}. We show the UNITsims-based HI power spectra with and without systematics as blue and black points, with $1\sigma$ error bars computed using Jackknife resampling (see Section \ref{sec:covariance}). The best-fit HI power spectra with and without systematics (see Equation \ref{eqn:pk_theory_and_sys}) are shown as accordingly colored solid lines, respectively. We find good agreement between $P_{\rm theory(,sys)}(k)$ and $P_{\rm HI(,sys)} (k)$, with differences between $1\sigma$ for $k < k_{\rm max}$ (see lower panel).  
Finally, we also plot the theoretical shot noise power spectrum $P_{\rm shot}$ as an orange solid line and the thermal-noise power spectrum $P_{\rm thermal}(k)$ as a magenta solid line. We see that $P_{\rm thermal}(k)$ rises rapidly and surpasses $P_{\rm HI,sys}$ already at $k = 0.1$ $h/{\rm Mpc}$. 

\autoref{fig:bias_comparison} represents the bias function $b(k)$ computed by dividing the HI-matter cross-power spectrum by the matter power spectrum, considering as tracers the halo catalogues, the HI galaxy catalogues and the HI Intensity maps. We extract the linear bias for each tracer by fitting the large scales $0.01 < k < 0.1$ $h/{\rm Mpc}$ to an horizontal line. All tracers are more clustered than dark matter, $b(k) > 1$. Both dark matter halo identification and the galaxy-halo connection provided by SAGE contribute to the bias of our HI Intensity maps. In the last step, we weight HI galaxies by its mass to create our HI Intensity maps. Since high mass galaxies are generally more biased there is no surprise to see an increase in the bias of the HI Intensity Maps with respect to the HI galaxy catalogue.

\section{Cosmological parameter estimation}
\label{section: cosmological_parameter_estimation}
We present in this section some forecasted parameter constraints we find for $\theta = $\{$\Omega_{\rm M}$, $\sigma_8$, $\sigma_8 b_1$\}. We rely on the Fisher information matrix formalism, which is a standard tool, quite computationally efficient, to predict uncertainties on cosmological parameters for a particular survey without available data. We evaluate the relative constraining power of the HI power spectrum and PDF for an SKA1-Mid HI Intensity Mapping observation modelling telescope beam and foreground-removal systematics for the first time.  

\subsection{Fisher forecast}
\label{sec:Fihser_analysis}

We compute the Fisher matrix as
\begin{equation}
    F_{ij} = \sum_{\alpha,\beta} \frac{\partial \rm{S_\alpha}}{\partial\theta_i}\Big|_{\rm fid} (C^{-1})_{\alpha \beta} \frac{\partial \rm{S_\beta}}{\partial\theta_j}\Big|_{\rm fid}\,,
\end{equation}
where $S_\alpha$ are the set of statistics, including the HI temperature power spectrum with $k_{\rm max} = 0.097$ $h/{\rm Mpc}$ and the HI temperature PDF, joining different smoothing scales ($R = 15.5, 24.8, 31.0$ ${\rm Mpc}/h$) and cutting the first and the two last quantiles. 
$C_{\alpha \beta}$ represents the data covariance matrix (see the details of its computation in \autoref{sec:covariance}) and $\partial \rm{S_\alpha}/\partial\theta_i$ are the summary statistic derivatives with respect to cosmological parameters, calculated in \autoref{sec:sensitivity_to_cosmological_parameters}.

The inverse of the Fisher matrix provides a lower bound of the parameter covariance matrix, under the assumption of a Gaussian likelihood and a linear response of the observables to variations in the cosmological parameters
\begin{equation}
    \sigma_{\theta_i} \geq \sqrt{\left[F^{-1}\right]_{ii}}\,.
\end{equation}

We checked that the distributions of PDF and $\rm{P(k)}$ bins are sufficiently close to Gaussian distributions (see \autoref{ap:test_gaussianity}). We also estimate the bias in the value of the cosmological parameters due to the difference between our simulated (UNITsims-based) and theoretically modelled statistics, which can be computed as
\begin{equation}
\label{eqn:fisher_bias}
    \delta \theta_i = \left[F^{-1}_{\rm mod}\right]_{ij}
    \sum_{\alpha,\beta}
    \frac{\partial S^{\rm mod}_\alpha}{\partial \theta_j}
    [C^{-1}]_{\alpha\beta}
    \Delta S_\beta\,,
\end{equation}
where $\Delta S_\beta = S_{\beta}^{\rm sim}-S_{\beta}^{\rm mod}$ is the data vector shift and $F^{-1}_{\rm mod}$ is the inverse Fisher matrix computed using modelled derivatives, which will be our default case. We consider the model to be reliable when the parameter biases are smaller than their corresponding uncertainties, i.e., $\delta \theta < \sigma_{\theta}$.

\subsection{Data Covariance}
\label{sec:covariance}

We estimate the covariance of the power spectrum and the one-point probability density function using jackknife resampling. HI Intensity Maps are divided into $N_{\rm box} = 5^3=125$ sub-boxes of length $L_{\rm subbox} = 200~{\rm Mpc}/h$. Our realizations are constructed by removing each one of those subboxes from the entire box. $N_{\rm box}$ is chosen to ensure both a sufficient sample size relative to the number of PDF bins used in this work, and that each sub-box encloses enough volume to allow reliable computation of the PDFs within it, which automatically makes the computation reliable for each realization.

Since we have four simulations we apply jackknife resampling to each individual box and then average the resulting covariance matrices over the four boxes. Note that we cannot combine subboxes from different boxes together to produce a shared covariance since we expect an error suppression for the power spectrum due to the fixed-and-paired technique applied to these simulations. 

The covariance matrix is estimated as: 

\begin{equation}
\label{eqn:Jack_c}
    C_{\rm ij}^{(m)} = \frac{ \left(N_{\rm box}-1\right)}{ N_{\rm box}  } \sum_{n=1}^{N_{\rm box}/4} \left(\mathbf{y}^{(n,m)}_{i}-\bar{\mathbf{y}}^{(m)}\right) \left(\mathbf{y}^{(n,m)}_{j}-\bar{\mathbf{y}}^{(m)}\right)
    \,, \hspace{2mm} C_{\rm ij} = \frac{1}{4} \sum_{m=1}^4 C_{\rm ij}^{(m)} \,,
\end{equation}
where $\mathbf{y}_{i}^{(n,m)}$ is the $i$-th element of the summary statistic computed on a realization where subbox $n$ is removed from box $m$, and $\bar{\mathbf{y}}^{(m)}$ is the mean of the summary statistic over all subboxes inside a box $m$. 

In our Fisher forecast, our data vector $\mathbf{y}_{i}^{(n,m)}$ represents either the power spectrum or its combination with the PDFs smoothed at a particular scale $R$. The power spectrum has linearly-spaced bins with $\Delta k = 6.3\times 10^{-3}$ $h/{\rm Mpc}$ and $k<k_{\rm max} = 0.097 $ $h/{\rm Mpc}$, with a total of $15$ bins. Note we could in principle use an analytical Gaussian covariance estimate for the HI power spectrum since we are considering large scales, but this estimate is not accurate when including observational systematics, which already have an impact in the power spectrum at large scales.

The PDF array is constructed for the smoothing scales $R = 15.5$, $24.8$, and $31.0$ ${\rm Mpc}/h$, using 20 of 22 quantiles per scale, excluding the first and the last quantiles to avoid very non-linear effects. Note that the bin centers for each smoothing scale should differ slightly.

\begin{figure}[htbp]
    \centering
    \begin{subfigure}[t]{0.43\textwidth}
        \centering
        \includegraphics[width=\textwidth]{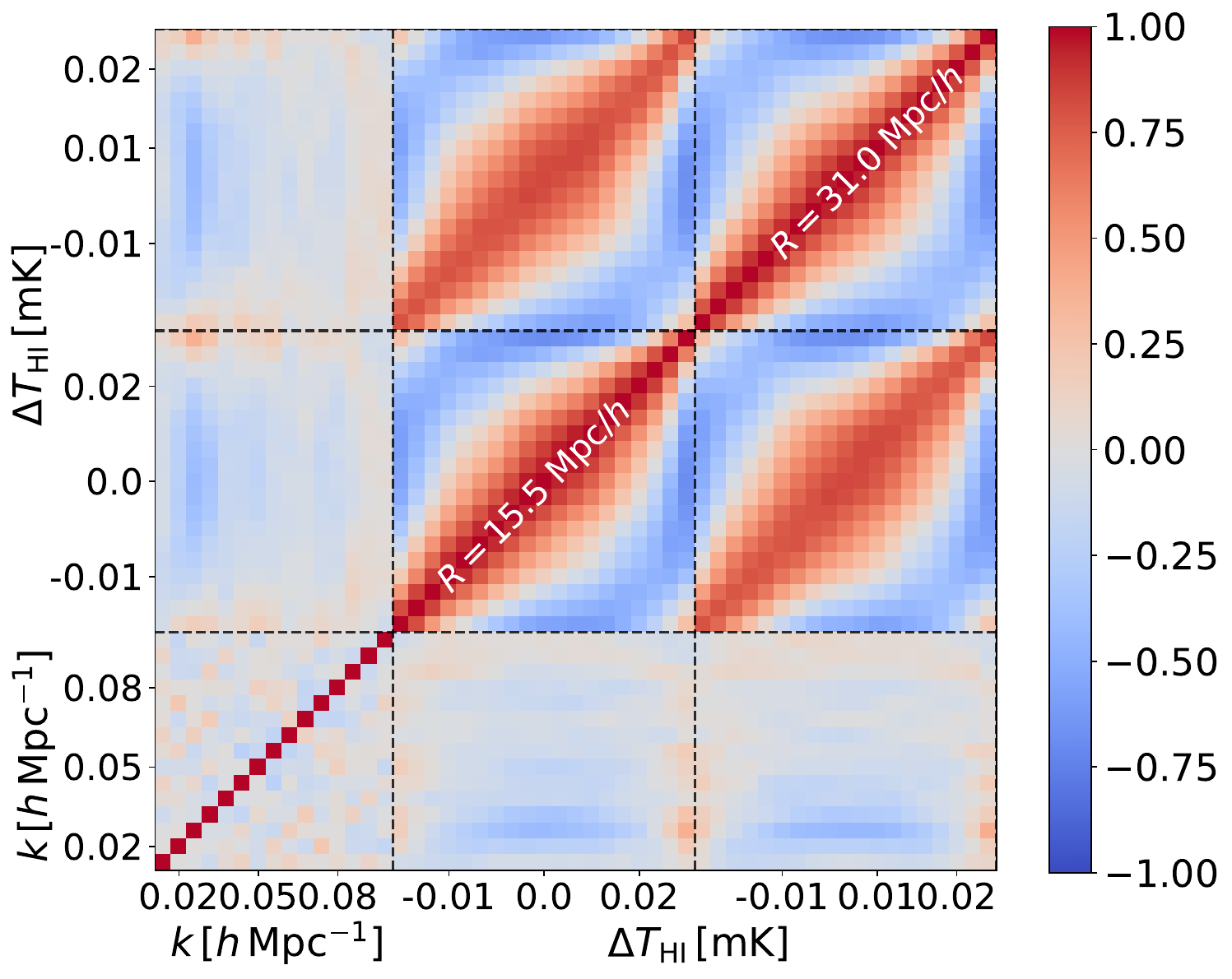}
        \caption{Correlation matrix}
        \label{fig:correlation}
    \end{subfigure}
    \hfill
    \begin{subfigure}[t]{0.5\textwidth}
        \centering
        \includegraphics[width=\textwidth]{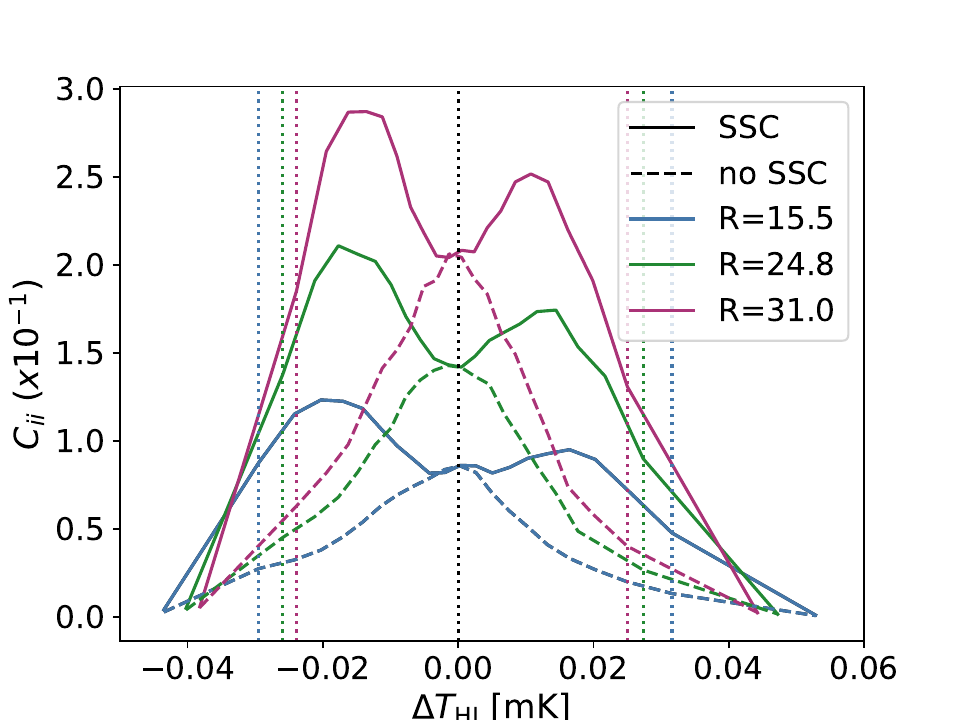}
        \caption{Covariance diagonal amplitudes}
        \label{fig:amplitudes}
    \end{subfigure}
    \caption{\textbf{Left: } Jackknife correlation matrix (removing SSC, equation \ref{eqn:super_sample}) for the power spectrum and PDFs smoothed at $R = 15.5$ and $31.0$ ${\rm Mpc}/h$ considering telescope beam ($R_{\rm beam} = 38.45$ ${\rm Mpc}/h$) and foreground-removal ($k_{\rm FG} = 3.64 \times 10^{-3}$ $h/{\rm Mpc}$) systematics. \textbf{Right: } Covariance diagonal amplitudes for all previous smoothing scales and including $R = 24.8$ Mpc/$h$: solid (dashed) lines consider super-sample covariance (removal). We keep PDF bins between colored dotted vertical lines. } 
    \label{fig:cov_matrix}
\end{figure}

We note that each subbox has its own mean overdensity. To approximately remove the impact of super-sample covariance (due to our unrealistically small subboxes) we evaluate our statistic for each subbox as
\begin{equation}
\label{eqn:super_sample}
    \mathcal{P}_i = \mathcal{P}_i\left(\frac{\Delta T_{\rm HI} +1}{\langle \Delta T_{\rm HI} \rangle_i+1} -1\right) \, ,
\end{equation}
where $\langle \Delta T_{\rm HI} \rangle_i$ is the mean temperature of subbox $i$. This effectively reduces the covariance amplitude arising from an effective shift in the mean and thus PDF peak location, as illustrated in \autoref{fig:amplitudes}. Although covariance amplitudes are significantly modified, the effect on Fisher forecasts is more modest with reductions up to 30 percent in parameter uncertainties. We will include the correction given by \autoref{eqn:super_sample} to compute covariances in the following sections unless otherwise is specified.

Given that we use an empirically estimated covariance matrix, its inverse is multiplied by the Kaufman-Hartlap factor \citep{kaufman1967,Hartlap}
\begin{equation}
    h_f = \frac{N_{\rm box}-p-2}{N_{\rm box}-1},
\end{equation}
where $p$ represents the total number of bins. This factor corrects for the bias of the precision matrix and it is close to unity for both statistics. 

In \autoref{fig:correlation} we present the correlation matrix for the power spectrum and PDFs at the two different smoothing scales. The symmetric matrix consists of six independent blocks: three auto-correlation blocks along the diagonal for the power spectrum and each PDF smoothing scale, one cross-correlation blocks between PDF scales and two cross-correlation blocks between the PDF and the power spectrum. We see positive correlations appearing near the PDF diagonals indicating positive correlations of neighbouring densities. Negative correlations appear farther from the diagonals, in particular between under- and overdensities. The strong correlations around the diagonals come from the fact that we use overlapping cells, which reduces the sampling error at the price of increasing the strength of non-diagonal terms \citep{Uhlemann2023PDFcovariance}. Notably, the cross-correlation blocks show strong correlation since densities in larger spheres inevitably include some of the density contained within smaller ones. In contrast, the off-diagonal elements of the auto-power spectrum show weaker correlations. We also observe a mild anti-correlation between large-scale power spectrum modes and the central region of the PDF, as expected as an increased variance leads to a lower peak height.

In \autoref{fig:amplitudes} we represent the amplitudes of the covariance diagonals with and without subtracting super-sample covariance as dashed and solid lines, respectively (see \autoref{eqn:Jack_c} and \autoref{eqn:super_sample}). We see that the absolute value of the covariance amplitude increases at higher $R$ slightly more than the square of the signal does as fewer independent cells are included in the boxes. Solid lines show two peaks in the amplitude at both sides of the mean density $\Delta T_{\rm HI} = 0$, with the highest one in underdensities due to the non-Gaussianity of the field. We also note that the peak position approaches  $\Delta T_{\rm HI} = 0$ at increasing $R$. Dashed lines show only one peak near 0 overdensities and show smaller amplitudes as solid lines specially at mild under- and over-densities.

\subsection{Sensitivity to cosmological parameters}
\label{sec:sensitivity_to_cosmological_parameters}
\begin{figure}[htbp]
    \centering
    \begin{subfigure}[t]{0.48\textwidth}
        \centering
        \includegraphics[width=\textwidth]{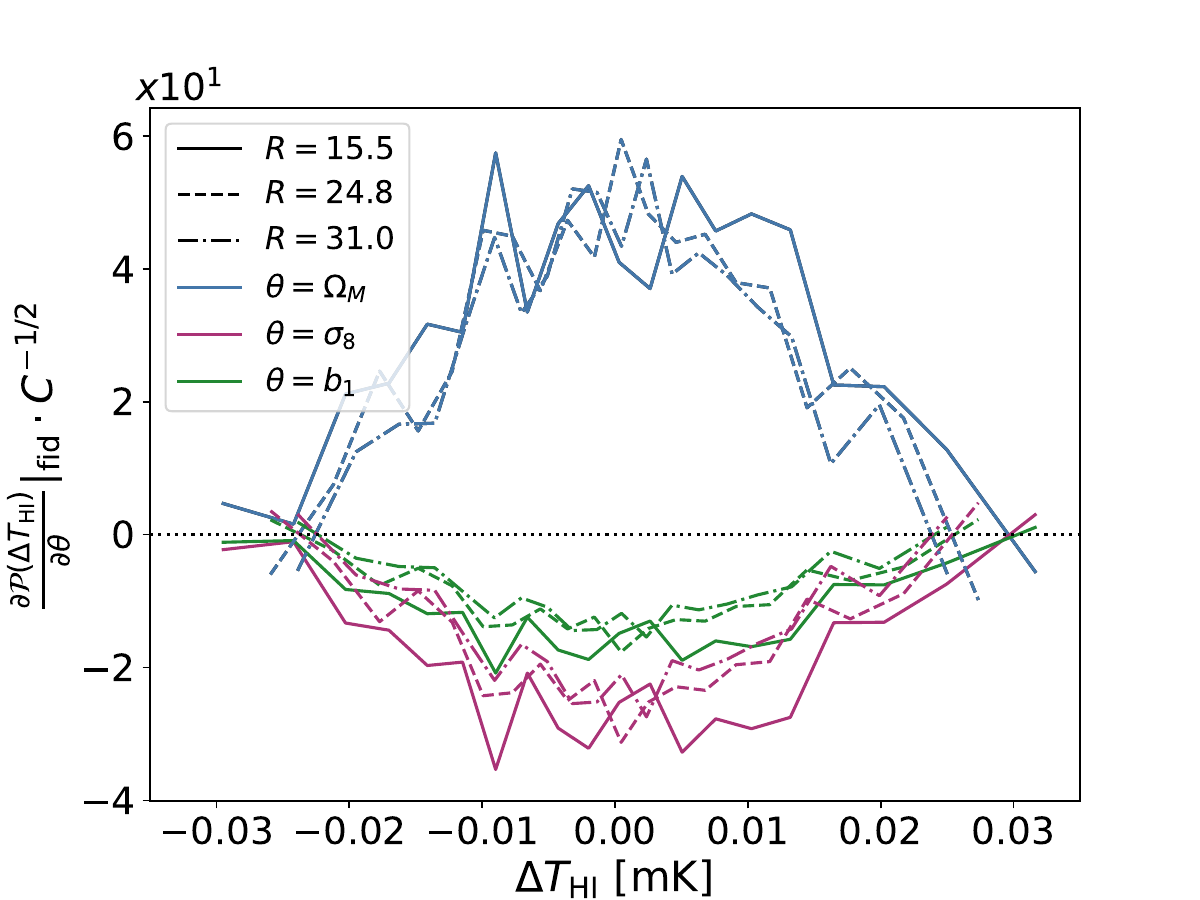}
        \caption{$\mathcal P(\Delta T_{\rm HI})$ for $R = 15.5, 24.8, 31.0{\rm Mpc}/h$ (solid, dashed, dotted)}
        \label{fig:PDFderivatives}
    \end{subfigure}
    \hfill
    \begin{subfigure}[t]{0.48\textwidth}
        \centering
        \includegraphics[width=\textwidth]{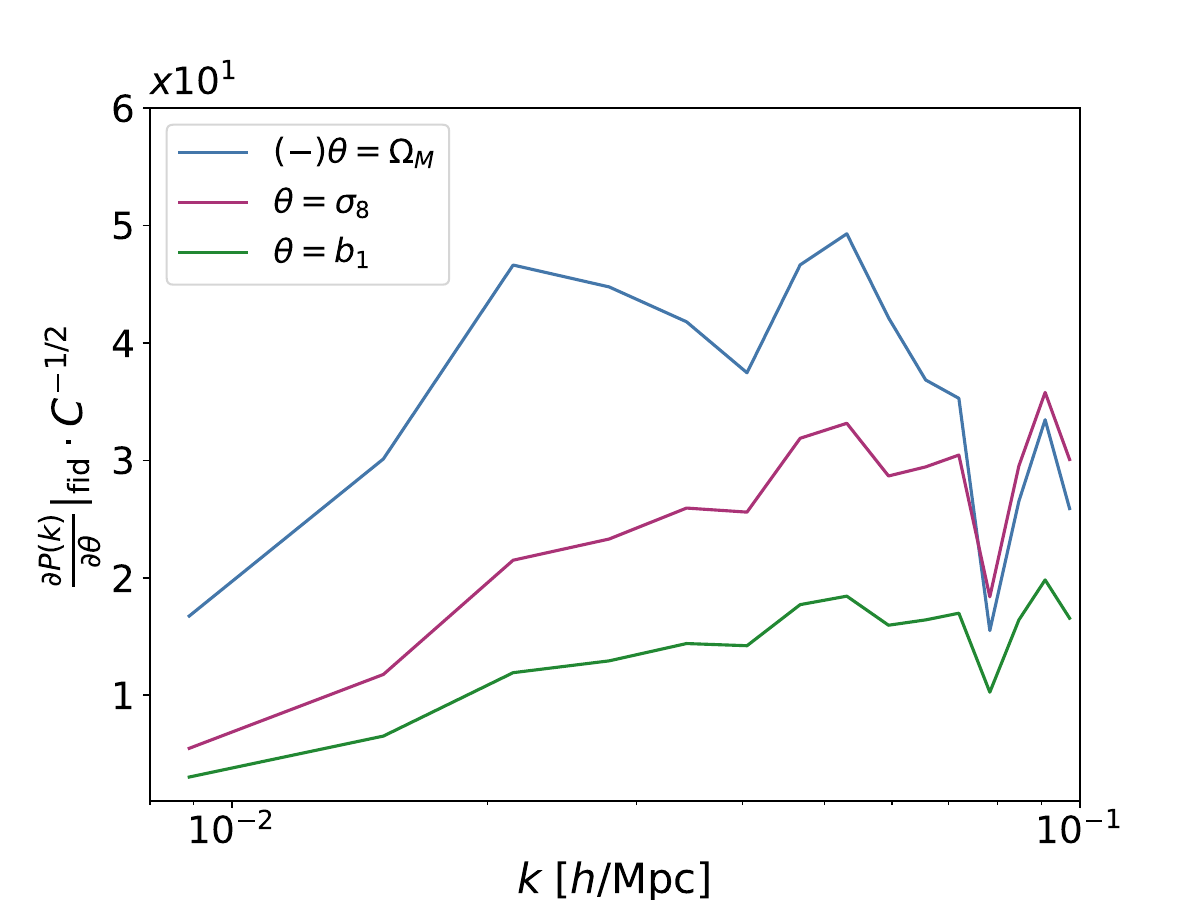}
        \caption{$P(k)$}
        \label{fig:Pkderivatives}
    \end{subfigure}
    \caption{Derivatives of the summary statistics  with respect to the cosmological parameters $\theta = \{\Omega_{\rm M},\sigma_8,b_1\}$, multiplied by the matrix-square root of the inverse covariance matrix.}
    \label{fig:derivatives}
\end{figure}
In this subsection, we analyze how our summary statistics depend on cosmological parameters. In \autoref{sec:joint_PDF} we provided theoretical predictions for $P_{\rm theory,sys}(k)$ and $\mathcal P(\delta_{\rm HI})$, where we assume in both cases a particular cosmology. To analyze the cosmology dependence of our summary statistics we perform cosmology shifts varying one parameter at a time, and compute the corresponding derivatives. 
We approximate the data vector derivatives with respect to all cosmological parameters with finite differences around a fiducial cosmology
\begin{equation}
    \frac{\partial S_\alpha} {\partial\theta_i}\Big|_{\rm fid} \approx \frac{ S_\alpha(\mathbf{\theta_{\rm fid}^i} + \mathbf{\Delta\theta_i}) -  S_\alpha(\mathbf{\theta_{\rm fid}^i} - \mathbf{\Delta\theta_i)}}{2\Delta\theta_i}\,,
\end{equation}
where $\mathbf{\theta_{\rm{fid}}}= \{\Omega_{\rm M},\Omega_b,\sigma_8, n_s,h,b_1 \}$ is the set of cosmological and linear bias parameters evaluated at the fiducial cosmology (UNITsims cosmology). We vary one parameter at a time and we select a step size of 3\%: $\mathbf{\Delta\theta_i} = 0.03 \theta_{\rm fid}^i$ at the parameter of interest and 0 otherwise. We checked the derivatives are consistent considering several step sizes between 3 to 20 percent of the fiducial value.  

\autoref{fig:PDFderivatives} shows the PDF derivatives with respect to cosmic parameters $\theta = \{\Omega_{\rm M},\sigma_8,b_1\}$ for three smoothing scales $R = 15.5, 24.8, 31.0$ ${\rm Mpc}/h$, multiplied by the matrix square root of the inverse covariance to estimate the contribution of each quantile bin to the constraining power for each cosmological parameter. The noise is coming entirely from the data covariance matrix as the derivatives are obtained analytically, and it is correlated through all PDF derivatives at a fixed scale (e.g. see the bumps for solid lines at $\Delta T_{\rm HI} = -0.01$ mK). Although PDF derivatives are also correlated at a fixed scale, which will allow to constrain only one parameter at a time, we are able to detect slightly different degeneracy directions when considering different smoothing scales. We calculate those by considering the scalar products of the Fisher matrix eigenvectors between different scales. We checked that those differences remain regardless of including or excluding the covariance matrix, which is the only source of noise. 
Finally, the PDF is most constraining for $\Omega_{\rm M}$ and $\sigma_8$, while changes on $b_1$ lead to signatures similar to $\sigma_8$, as they all change the variance and skewness.  

\autoref{fig:Pkderivatives} shows the power spectrum derivatives multiplied by the inverse of the square root of the covariance. Note we multiply the derivative with respect to $\Omega_{\rm M}$ by $-1$ for representation purposes. Changes on $b_1$ and $\sigma_8$ only differ by a constant term, as the power spectrum depends linearly on both parameters.

\subsection{Results}
\label{sec: results_fisher}

\begin{center}
    \begin{figure}[h]
        \centering
    	\includegraphics[scale=0.5]{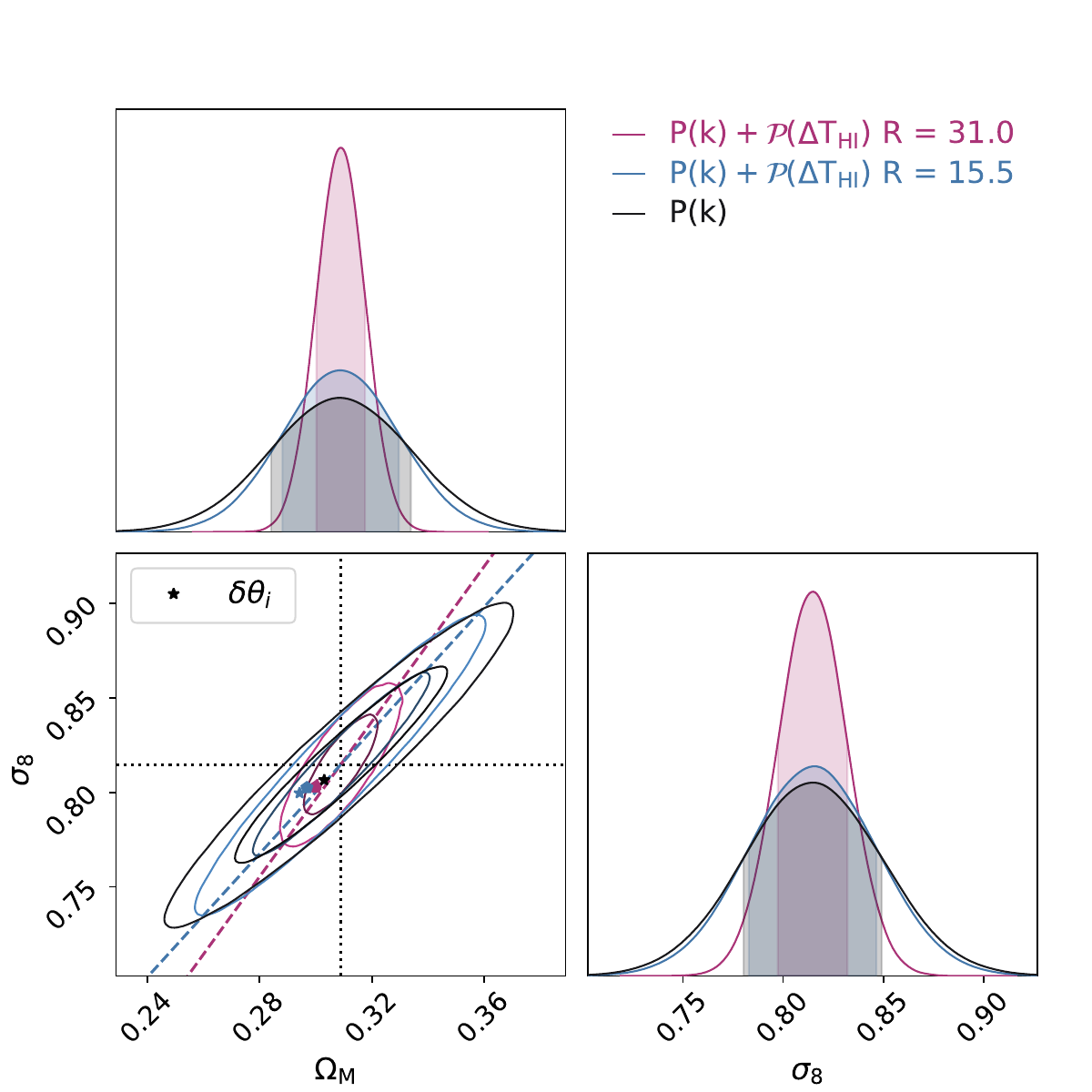}
    	\caption{\{$\Omega_{\rm M}$, $\sigma_8$\} Fisher forecast where we show 1$\sigma$ and 2$\sigma$ contours for the HI power spectrum alone $\rm{P(k)}$ (black) and its combination with the PDF  ($\rm{P(k)}$ + $\mathcal{P}(\Delta T_{\rm HI})$ at R = 15.5,31.0 ${\rm Mpc}/h$) in blue and red  ellipses, respectively. Colored stars represent the parameter bias recovered due to differences between UNIT-based and model statistics, which are colored accordingly. Colored pentagons indicate the parameter bias calculated after updating our model PDF skewness to match the theoretical skewness we get from integrating the tree-level bispectrum with systematics. Dashed lines represent the degeneracy directions of PDF-only forecasts.}
        \label{fig:Contour_plot_PDF_PDF}
    \end{figure}
\end{center}

In \autoref{fig:Contour_plot_PDF_PDF} we show forecasted constraints on $\Omega_{\rm M}$ and $\sigma_8$ (at fixed bias $b_1$) considering the HI power spectrum $\rm{P(k)}$ and its combination with the HI matter PDF at one scale $\rm{P(k)}$ + $\mathcal{P}(\Delta T_{\rm HI})$ at $R=15.5$ Mpc/$h$ as well as $R=31.0$ Mpc/$h$ ${\rm Mpc}/h$. $P(k)$ constraints are represented in black, where we see a degeneracy between both parameters. The correlation factor is $r_{\rm Pk} = 0.964$. By adding the PDF at $R = 31.0$ ${\rm Mpc}/h$ we  reduce the degeneracy  to $r_{\rm P(k)+\mathcal{P}(\Delta T_{\rm HI})} = $ $0.852$ and thus reduce uncertainties by 64 and 49 percent in $\Omega_{\rm M}$ and $\sigma_8$, respectively. Considering a smaller smoothing scale for the PDF, $R = 15.5$ ${\rm Mpc}/h$,  $\{\Omega_{\rm M}$,$\sigma_8\}$ uncertainties are reduced by 9 and 1 percent, respectively. Although it might be intuitively expected that a smaller smoothing scale would yield tighter parameter constraints (as smaller scales are more non-Gaussian), the effectiveness of the combination of a single-scale PDF with the power spectrum primarily depends on the relative orientation of their degeneracy directions. In this particular case, increasing the smoothing scale leads to a progressive misalignment between the degeneracy direction of the PDF and that of the power spectrum, which helps to enhance their complementarity. Note that large scales are more sensitive to the total amount of matter, $\Omega_{\rm M}$, while smaller scales tend to be more sensitive to matter fluctuations, $\sigma_8$. 

    \begin{figure}[h]
        \centering
    	\includegraphics[scale=0.6]{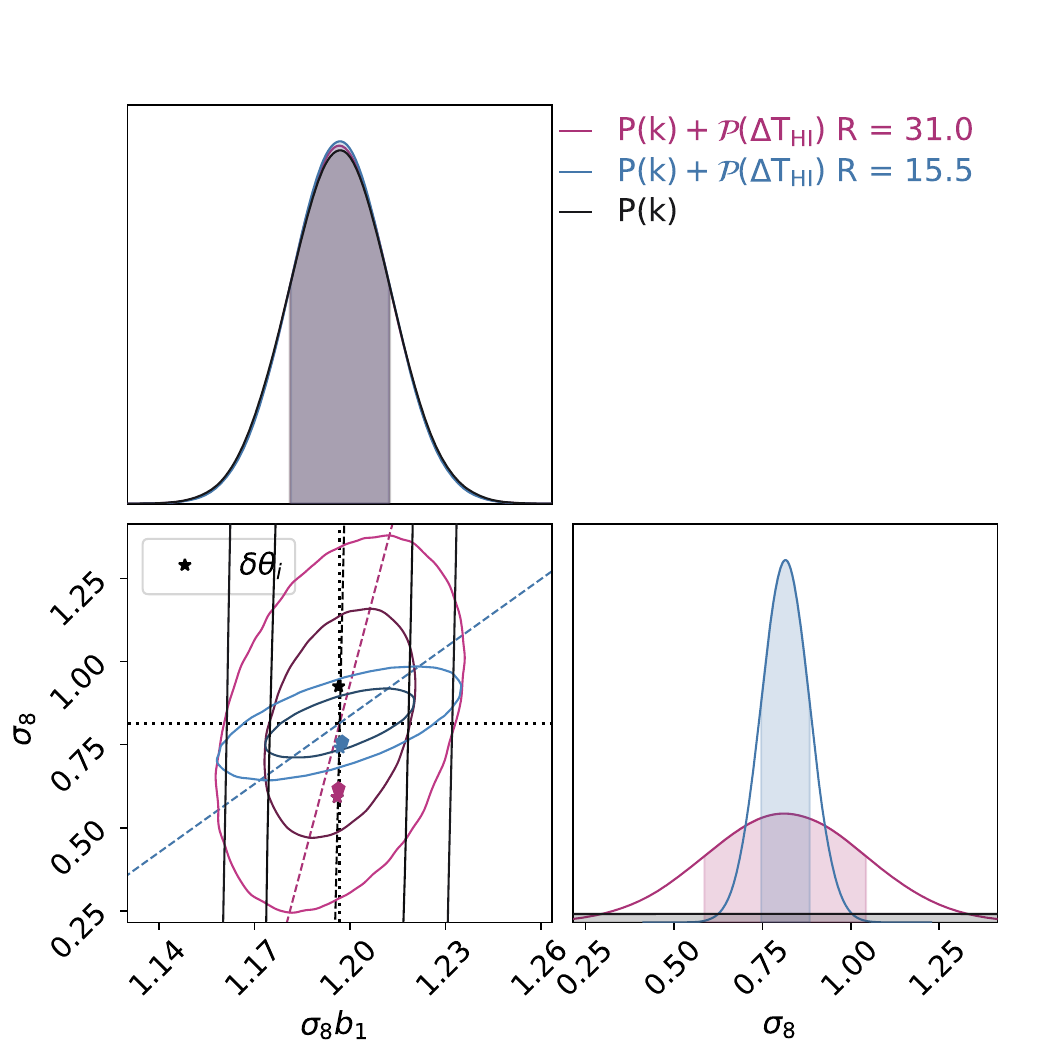}
    	\caption{Same as \autoref{fig:Contour_plot_PDF_PDF} considering \{$\sigma_8$, $\sigma_8 b_1$\} as free parameters.}
        \label{fig:Contour_plot_s8_b1}
    \end{figure}
In \autoref{fig:Contour_plot_s8_b1} we forecast constraints for $\sigma_8$ and $\sigma_8 b_1$ together using the same combined statistics. We can see again that adding the PDF helps reducing uncertainties. In this configuration, degeneracy contours favour smaller smoothing scales for the PDF, which seem promising breaking the degeneracy between $\sigma_8$ and $b_1$ as expected from the increase in non-Gaussianity signal on smaller scales. This behaviour can be understood in terms of the reduced skewness. For the matter field, the reduced skewness,
$S_{3,\rm m}=\langle\delta_{\rm m}^3\rangle /\sigma_{\rm m}^4$
is approximately independent of the fluctuation amplitude, remaining nearly constant when expressed as a function of the variance. In contrast, for a linearly biased tracer, $\delta_{\rm HI}=b_1\delta_{\rm m}$, the reduced skewness is
$S_{3,\rm HI}=\langle\delta_{\rm HI}^3\rangle /\sigma_{\rm HI}^4
\simeq S_{3,\rm m}/b_1$ \citep{FryGaztanaga1993,Uhlemann_2018a}.
Therefore, a non-zero skewness in the PDF provides information that is primarily sensitive to the linear bias while only weakly depending on $\sigma_8$. Combined with the two-point statistics, which constrain the product $\sigma_8 b_1$, this additional information helps disentangle the two parameters. Again, we show the cosmological parameter biases for the simplified PDF systematics model as colored stars and the Edgeworth-modified systematics model as pentagons, respectively.

When extending the parameter space to include $\Omega_{\rm M}$ in addition to $\{\sigma_8,\sigma_8 b_1\}$, the parameter uncertainties increase compared to the two-parameter analysis. For $R=15.5\,\mathrm{Mpc}/h$, the uncertainty on $\sigma_8 b_1$ increases by approximately a factor of three, whereas the uncertainty on $\sigma_8$ remains nearly unchanged. In contrast, for $R=31.0\,\mathrm{Mpc}/h$, the uncertainties on both $\sigma_8 b_1$ and $\sigma_8$ increase by approximately a factor of two.

The covariance matrix used in this analysis is adapted to the simulation volume of $1$ $\left(\rm{Gpc}/h\right)^3$.
We represent the parameter bias we recover in cosmological parameters (see \autoref{eqn:fisher_bias}) for statistics as colored stars and record the values in Table~\ref{tab:Fisher_parameter_bias}. For the simplistic theoretical systematics treatment, we obtain relative parameter biases of $0.5\sigma$ to $1\sigma$. By including an Edgeworth-based systematics correction~\eqref{eqn:DM_PDF_sys_Edgeworth_2} in our modelled dark matter PDFs to match the theoretical skewness from integrating the tree-level bispectrum weighted with systematics (see Appendix \ref{sec:skewness} for the details), the parameter biases are slighly reduced. These results are shown as the corresponding coloured pentagons.

SKA1-MID surveys will be able to map much larger volumes than the simulations that were available for the data covariance matrix. To forecast parameter uncertainties for SKA1-MID \citep[as done in][]{Majumdar2026}, the  effective survey volume would be
\begin{equation}
    V_{\rm s} = S_{\rm area} \frac{c}{H_0} \int_{z_{\rm min}}^{z_{\rm max}} dz \frac{r(z)^2}{E(z)} = 16.3 \hspace{2mm} \left(\rm{Gpc}/h\right)^3 \, ,
\end{equation}
where $S_{\rm area} = f_{\rm sky} \times 41252.96 \left(\frac{\pi}{180}\right)^2$ steradians with a sky coverage fraction of $f_{\rm sky} = 0.3$, $z_{\rm min} = 1.1$ and $z_{\rm max} = 1.5$ as well as a comoving distance $r(z)$ and Hubble parameter $H(z) = H_0 E(z)$ at the fiducial cosmology. As parameter uncertainties scale with the survey volume as $\sigma (\theta) \propto V^{-1/2}$ while the Fisher bias is approximately volume-independent, the relative impact of the bias, quantified by 
$\delta \theta/\sigma_\theta$, would be amplified by the same factor. We leave for future work the consideration of bigger simulations to refine our modelled PDFs. 

\begin{table}[]
    \centering
    \begin{tabular}{c|c|c|c|c}
      & \multicolumn{2}{c|}{$\delta \Omega_{\rm M} / \sigma_{\Omega_{\rm M}},\delta \sigma_8 / \sigma_{\sigma_8}$} & \multicolumn{2}{c}{$\delta (\sigma_8 b_1) / \sigma_{(\sigma_8 b_1)},\delta \sigma_8 / \sigma_{\sigma_8}$}\\ \hline
      $R$ [Mpc$/h$] & $15.5$ & $31.0$ & $15.5$ & $31.0$\\\hline
    $S_3(P(k)S^2_{\rm eff}(k))$ & $-0.71,-0.48$ & $-0.96,-0.63$ & $0.01, -1.08$ & $-0.05, -0.99$ \\\hline
    $S_3^{\rm tree,B}$ & $-0.59$, $-0.39$ & $-1.03$, $-0.69$ & $0.05$, $-0.83$ & $-0.02$, $-0.87$\\ 
    \end{tabular}
    \caption{Relative Fisher parameter bias obtained for the combination of the HI power spectrum and the single-scale HI PDF for the smallest and largest scale for the simplified systematics modelling  $S_3(P(k)S^2_{\rm eff}(k))$ and the improved systematics modelling using the skewness value from the tree-order bispectrum $S_3^{\rm tree, B}$.}
    \label{tab:Fisher_parameter_bias}
\end{table}

\section{Conclusion}
\label{sec:concl}

In this work we determined how radio survey systematics such as the telescope beam effect, foreground-removal effect and thermal noise affect the one-point Probability Density Function (PDF) of simulated Intensity maps of neutral hydrogen (HI). We provided a theoretical model for the HI PDF including systematics, validated it with simulated intensity maps and demonstrated how it can complement the power spectrum by constraining cosmological parameters and the linear bias.

We used the public code   \href{https://github.com/OliverFHD/CosMomentum}{\textsc{CosMomentum}} \citep{Friedrich2025jointPDF} to model the HI PDF  at three mildly nonlinear smoothing scales. \textsc{CosMomentum} predicts the tracer PDF with large-deviation theory and a conditional PDF of tracer given matter densities described by a conditional mean (tracer bias) and variance (tracer stochasticity). We showed that a quadratic Eulerian bias model and a non-Poissonian quadratic stochasticity model can successfully model the SKA-like simulated Intensity Maps from the UNIT simulations at  mildly nonlinear scales from 15-30 Mpc/$h$.  
We incorporated observational systematics by modifying the rate function where linear and non linear variances are computed now integrating model power spectra together with the corresponding systematic kernels. We compared our theoretical and simulated HI PDF in \autoref{fig:PDF_CosMomentum_no_sys_and_sys}, considering a jackknife-based covariance finding percent-level agreement for both systematic and systematic-free tests. In the absence of systematics,  the smallest smoothing scale of $R = 15.5$ Mpc$/h$ shows discrepancies up to 4 percent indicating the need for a higher order bias description \citep[see][]{Gould2025}.  

To estimate the expected constraining power of the HI PDF as complement to the power spectrum, we performed a Fisher forecast.
We model the mildly nonlinear HI power spectrum using \textsc{hmcode-2020}, implemented in \textsc{camb} with a linear HI bias and shot noise in form of white noise. Observational systematics are included through a $k-$dependent power spectrum modulation, $S_{\rm eff}(k)$, obtained from integrating the beam and foreground kernels. We demonstrated the effectiveness of our simplistic power spectrum model in \autoref{fig:power_and_bias}, using a jackknife-based covariance.

We used our theoretical models for the HI PDF and power spectrum to predict the signal to noise for cosmological and bias parameters $\Omega_{\rm M}, \sigma_8$ and $b_1$ shown in \autoref{fig:derivatives}. With those we performed a Fisher analysis to compare the constraining power of the power spectrum alone with that obtained by combining it with the HI PDF at different smoothing scales. When combining the HI PDF at a single scale with the power spectrum in our forecasts, the most misaligned degeneracy direction with respect to the power spectrum is giving the best constraints. We provided two examples constraining together $\Omega_{\rm M}$ and $\sigma_8$ and another forecast constraining $\sigma_8$ and $\sigma_8 b_1$. In the first example in \autoref{fig:Contour_plot_PDF_PDF} we showed that the degeneracy between $\Omega_{\rm M}$ and $\sigma_8$ can be reduced when including the PDF at the highest smoothing scale, since is the one with the degeneracy direction the most misaligned with the power spectrum. In the second example represented in \autoref{fig:Contour_plot_s8_b1}, the PDF at the smallest smoothing scale—corresponding to the case that contains the most non-Gaussian information—provided the strongest improvement in breaking the degeneracy between $\sigma_8$ and $b_1$. In both cases we can noticeably tighten the constraints on $\Omega_{\rm M}$, $\sigma_8$ and $b_1$ and help breaking the $\sigma_8-b_1$ degeneracy. The parameter bias due to differences between simulated and modelled statistics is under $1\sigma$ for all forecasts without volume-rescaling our covariance. If we perform an Edgeworth-based systematics correction to the matter PDF such that its skewness matches the one computed theoretically in Appendix \ref{sec:skewness}, parameter biases are slightly reduced, but higher volumes probed by the SKA survey might require a refined model for the combination of the power spectrum and PDF and its covariance. 

We leave for future work a more detailed understanding of the skewness rescaling in the presence of systematics, the marginalisation of shot noise parameters as well as the use of larger volume or higher number of simulations to enable a more accurate and precise modelling of the HI PDF signal and covariance. Another promising extension is the development of angular one-point statistics \citep[see e.g.][]{Uhlemann_2018,Friedrich_2021a,Friedrich2025jointPDF}, computed from thin cylindrical volumes along the line of sight, which would naturally exploit the excellent radial resolution of HI intensity mapping surveys and establish a closer connection with one-point statistics used in weak-lensing analyses.

\section*{Acknowledgments}

BVG and CU were supported by the European Union (ERC StG, LSS\_BeyondAverage, 101075919). We thank Gustavo Yepes and Alexander Knebe for providing the dark matter and galaxy simulations, respectively. We thank Beth McCarthy Gould, Lina Castiblanco and Oliver Friedrich for discussions. This research was supported by the Munich Institute for Astro-, Particle and BioPhysics (MIAPbP) which is funded by the Deutsche Forschungsgemeinschaft (DFG, German Research Foundation) under Germany´s Excellence Strategy – EXC-2094 – 390783311. The CosMomentum code underpinning this work resulted from a collaboration that was supported by the Deutsche Forschungsgemeinschaft (DFG, German Research Foundation) via the PaNaMO project (project number 528803978) and we thank Oliver Friedrich for publicly releasing and maintaining it. 
\bibliographystyle{apsrev4-1}

\bibliography{oja_template}

\begin{appendix}

\section{Impact of systematic kernels on one-point statistics}
\label{ap:ap1}

In this appendix we discuss how the telescope beam and the foreground removal impact the shape of the PDF. In particular we focus on the modelling of their effects on the variance, $\sigma^2$, and the reduced skewness, $S_{\rm 3}$. We also provide a comparison between the theoretical and simulated dark matter PDFs, with and without including systematics. Finally, we investigate how systematics modify the effective volume, which, in the absence of such effects, is determined only by the smoothing scale $R$ used to estimate tracer stochasticity.

\subsection{Variance}
\label{sec: variance_computations}
In this appendix we compare the UNIT smoothed variance with systematics with the variance obtained by integrating the theoretical matter power spectrum without systematics, $P_{\rm m}(k)$ (see section \ref{sec:power_spectrum_theory}), convolved with both theoretical kernels $f (k_\parallel)$ and $B (k_\perp)$ and a window function, for several smoothing scales $R$. We perform a similar comparison without considering systematics and considering the HI field as well. 
\begin{equation}
\label{eqn:variance_nosys}
    \sigma_{\rm HI (m)}^2 (R) = \frac{1}{2 \pi^2} \int_{\rm k_{\rm F}}^{\rm k_{\rm nyq}} d k \hspace{1mm} k^2 P_{\rm theory (m)} (k) W^2_{\rm TH} (kR) \,,
\end{equation}

\begin{center}
    \begin{figure}[!t]
        \centering
    	\includegraphics[scale=0.35]{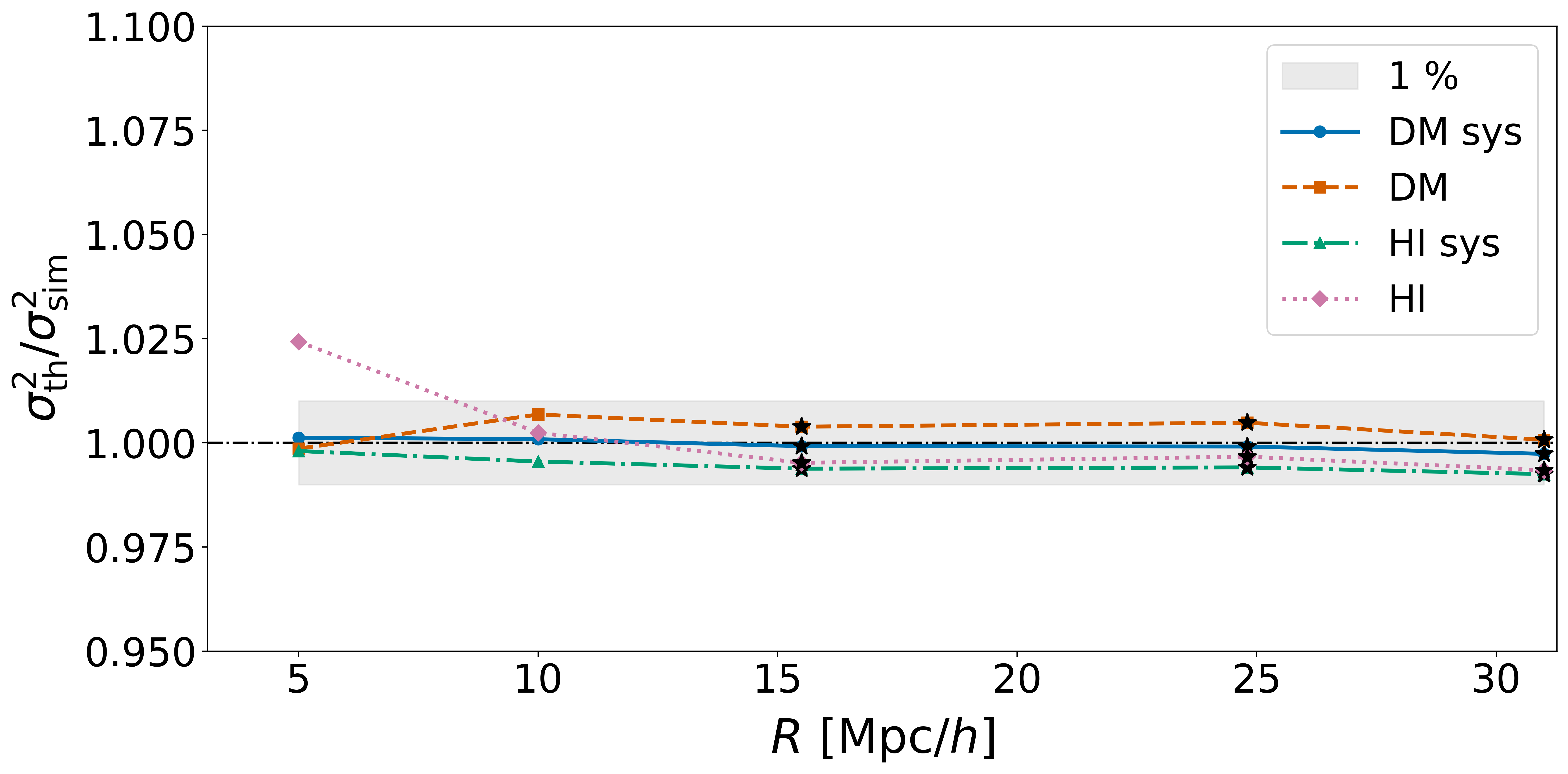}
    	\caption{Ratio between the smoothed variance calculated by integrating the theoretical power spectrum $P(k)$ of both dark matter and HI catalogues with and without systematics (see \autoref{eqn:variance_nosys} and \autoref{eqn:variance_kernels}) and its corresponding variance calculated using UNIT simulations. The 1\% region is shown as a grey shaded area. Black star markers indicate the smoothing scales used in this work.}
    \label{fig:variance_comparison}    
    \end{figure}
\end{center}

\begin{equation}
\label{eqn:variance_kernels}
    \sigma_{\rm HI (m),sys}^2 (R, R_{\rm beam}, k_{\rm FG}) =  \frac{1}{2 \pi^2} \int_{\rm k_{\rm F}}^{\rm k_{\rm nyq}} d k \hspace{1mm} k^2 P_{\rm theory (m)} (k) W^2_{\rm TH} (kR) S_{\rm eff} (k; R_{\rm beam}, k_{\rm FG} ) \,,
\end{equation}
where $k_{\rm F} = 6.28 \times 10^{-3}$ $h/{\rm Mpc}$ is the fundamental mode, $k_{\rm nyq} = 3.14$ $h/{\rm Mpc}$ is the Nyquist mode and $S_{\rm eff}$ is given in \autoref{eqn:Seff_k}. Note that the integrand is highly suppressed by the telescope beam much before reaching $k_{\rm nyq}$. For the case of beam systematics, analytical results can be obtained by recognising that the angular integral can be obtained in terms of the Dawson or imaginary error function
\begin{equation}
    B_{\rm eff}^2(k)=\frac{1}{2}\int_{-1}^{+1} d\mu B(kR_{\rm beam}\sqrt{1-\mu^2})^2=\frac{\text{Dawson}(kR_{\rm beam})}{kR_{\rm beam}}=\frac{\sqrt{\pi}}{2}{\rm e}^{-(kR_{\rm beam})^2}\frac{{\rm erfi}(kR_{\rm beam})}{kR_{\rm beam}}\,.
\end{equation}

For the combination of beam and foreground cleaning systematics, we have
\begin{equation}
S_{\rm eff}^2(k) = 
    \frac{\sqrt{\pi}}{2}{\rm e}^{-(kR_{\rm beam})^2} \left[
    \frac{{\rm erfi}(kR_{\rm beam})}{kR_{\rm beam}}
    -
    2 \frac{k_{\rm FG} {\rm erf}[(\frac{k}{k_{\rm FG}} \sqrt{1 - k_{\rm FG}^2 R_{\rm beam}^2})}{k\sqrt{1 - k_{\rm FG}^2 R_{\rm beam}^2}} + \frac{k_{\rm FG} {\rm erf}[(\frac{k}{k_{\rm FG}} \sqrt{2 - k_{\rm FG}^2 R_{\rm beam}^2})]}{k
   \sqrt{2 - k_{\rm FG}^2 R_{\rm beam}^2}}
    \right] \,.
\end{equation}

\autoref{fig:variance_comparison} shows the variance ratios between theory and simulations for matter and HI, with and without systematics. We see a general agreement under $1 \%$, although deviations between the two curves at small smoothing radii ($R < 10$ ${\rm Mpc}/h$) may indicate the presence of non-linear effects. For our analysis, we adopt $R = 15.5, 24.8, 31.0$ ${\rm Mpc}/h$ smoothing scales. 

When modelling systematics in \textsc{CosMomentum}, we replace Equation~\ref{eqn:variance_nosys} with Equation~\ref{eqn:variance_kernels} when computing both the linear and non-linear variances. 

\begin{center}
    \begin{figure}[!t]
        \centering
    	\includegraphics[scale=0.5]{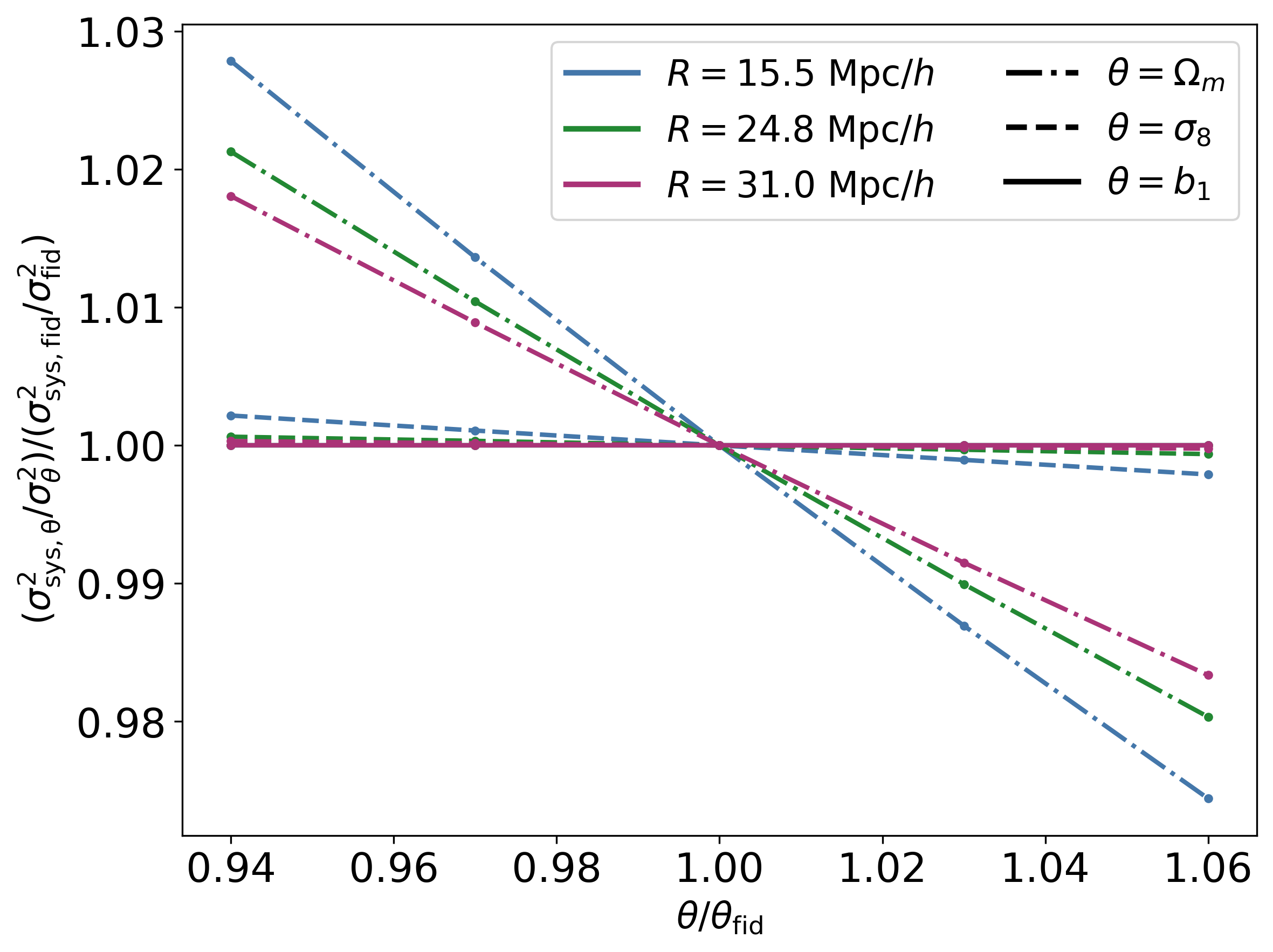}
    	\caption{Cosmological parameter $\theta=$\{$\Omega_{\rm M},\sigma_8,b_1$\} dependence between variance ratios with and without systematics as a function of $\theta/\theta_{\rm fid}$, for smoothing radius $R = 15.5, 24.8, 31.0$ ${\rm Mpc}/h$. }
    \label{fig:variance_ratio_cosmology}    
    \end{figure}
\end{center}

In \autoref{fig:variance_ratio_cosmology} we analyse the cosmology dependence of the theoretical variance ratios with and without systematics analysing variations up to 6 percent in $\theta=$\{$\Omega_{\rm M},\sigma_8,b_1$\}, for the three smoothing scales considered in this work. $\Omega_{\rm M}$ is the most affected parameter, with differences up to $2 - 3$ percent when $\theta/\theta_{\rm fid} = 1.06$. We see almost no dependence when varying $\sigma_8$ or $b_1$ since both are global parameters and the first term in equation \ref{eqn:pk_theory_and_sys} clearly dominates. We find for all parameters a linear evolution of the variance ratio with cosmology. Finally we see that differences increase slightly when the smoothing scale is decreasing. We take into account those differences in our theoretical model for the HI PDF to avoid possible bias when constraining cosmological parameters.

\subsection{Skewness}
\label{sec:skewness}
In addition to the change in the variances, the presence of the beam and foreground removal systematics cause an associated change in the reduced skewness defined as $S_3(R)=\langle\delta_R^3\rangle/\langle\delta_R^2\rangle^2$. If we consider a smoothed density $\delta_R$ obtained from spherically symmetric filter on a given scale $R$ with a Fourier kernel $W_{\rm TH}(kR)$, then at tree-order perturbation theory
\begin{align}
\label{eq:S3_sphere}
S_3(R)=\frac{\langle\delta_R^3\rangle}{\langle\delta_R^2\rangle^2} &\approx 6 \frac{\int \text{d}^3k_1 \text{d}^3k_2 P(k_1)P(k_2)W_{\rm TH}(k_1 R)W_{\rm TH}(k_2 R) W_{\rm TH}(|\vk_1+\vk_2|R)F_2(-\vk_1,-\vk_2)}{\left(\int \text{d}^3k_1  P(k_1)W_{\rm TH}(k_1 R)\right)^2}= \frac{34}{7} + \frac{d\log\sigma^2(R)}{d\log R}\,,
\end{align}
where the second-order perturbation kernel is \citep{Bernardeau2002review}
\begin{equation}
F_2(\vk_1,\vk_2)=\frac{5}{7}+\frac{\vk_1 \cdot \vk_2}{2}\left(\frac{1}{k_1^2}
+\frac{1}{k_2^2}\right)+
\frac{2}{7}\frac{(\vk_1 \cdot \vk_2)^2}{k_1^2 k_2^2}\,.
\end{equation}

This result is obtained by re-expressing the angular average of  $W_{\rm TH}(|\vk_1+\vk_2|R)F_2(\vk_1,\vk_2)$ in terms of a sum of products $W_{\rm TH}(k_1 R)W_{\rm TH}(k_2 R)$ and $W_{\rm TH}(k_1 R)k_2W_{\rm TH}'(k_2 R)$, where the derivative with respect to the argument of $W$ can be written as a derivative with respect to the radius $R$ \citep{Bernardeau94skewkurt}.
When considering survey systematics, we have a combined filter including spherical, angular and perpendicular components $W(\vk)=W_{\rm TH}(|\vk|R)B(\vk_\perp R_{\rm beam})f(k_\parallel/k_{\rm FG})$ and need to compute
\begin{equation}
\label{eq:S3_beam}
S_3(R,R_{\rm beam},k_{\rm FG})\approx 6 \frac{\int \text{d}^3k_1 \text{d}^3k_2 P(k_1)P(k_2)W(\vk_1)W(\vk_2) W(-\vk_1-\vk_2)F_2(-\vk_1,-\vk_2)}{\left(\int \text{d}^3k_1  P(k_1)W(\vk_1)^2\right)^2}\,.
\end{equation}

In the computation of our spherical top-hat PDF we included the impact of the angular and line-of-sight kernels  at the level of the power spectrum, essentially assuming the simplification of equation~\eqref{eq:S3_sphere}, which is known to be modified for filters of more complex shapes. For example, for general spherically symmetric filters the correction can be determined using equation 25 in \cite{Bernardeau2016} and for cylindrical filters as described in Appendix A3 in \cite{Uhlemann_2018}. When numerically comparing the reduced skewness from the beam-consideration following equation~\eqref{eq:S3_beam} we find that the approximation overestimates the skewness in particular for larger sphere radii, see Table~\ref{tab:skewness_app}. 
\begin{table}[]
    \centering
    \begin{tabular}{c||c||c|c|c||c}
             $R$ [Mpc/h] & $S_{\rm 3,th}(P(k)S^2_{\rm eff}(k))$ & $S_{\rm 3,th}^{\rm tree}(P(k)S^2_{\rm eff}(k))$  & $S_{\rm 3,th}^{\rm tree}(P(k)B^2_{\rm eff}(k))$  &  $S_{\rm 3,th}^{\rm tree, B}$ & $S_{\rm 3,sim}$\\\hline
        15.5  & $4.47$ & $4.40$ & $4.42$ &  $4.22$ & $4.34$\\ 
         24.8 & $4.16$ & $4.15$ & $4.20$ & $3.76$ & $3.82$\\ 
         31.0 & $3.98$ & $4.00$ & $4.06$ & $3.57$ & $3.58$ \\
    \end{tabular}
    \caption{Comparison of reduced skewness values for matter with systematics in the sphere-like approximation~\eqref{eq:S3_sphere} with a modified power spectrum including the beam and foregrounds at the nonlinear and tree level. For the beam only, we recompute the tree level results with a proper beam-consideration following equation~\eqref{eq:S3_beam}. The tree-level results including the proper beam effect approach the simulated values (last column) at larger smoothing scales where the tree-level approximation is effective.}
    \label{tab:skewness_app}
\end{table}
To approximately correct for the impact of the systematics on the reduced skewness, we implement an Edgeworth-style residual correction as follows
\begin{equation}
\label{eqn:DM_PDF_sys_Edgeworth_2}
    \mathcal P(\delta_{\rm m,sys}) \rightarrow \frac{1}{\sqrt{2\pi\sigma_m^2}}\exp\left(-\frac{\delta_m^2}{2\sigma_m^2}\right)\left[1+\frac{S_3\sigma_{\rm m}}{6} \rm{He}_3\left(\frac{\delta_m}{\sigma_m}\right)+\frac{S_4\sigma_m^2}{24}\rm{He}_4\left(\frac{\delta_m}{\sigma_m}\right) + \frac{(S_3\sigma_m)^2}{72}\rm{He}_6\left(\frac{\delta_m}{\sigma_m}\right)\right]\,,
\end{equation}
where $S_3=S_{3,\rm th}^{\rm tree,B}$ is the reduced skewness at tree level from Table~\ref{tab:skewness_app},  $S_4=\langle\delta^4\rangle_c/\sigma^6$ is kept fixed from the approximate dark matter PDF with systematics from \texttt{CosMomentum}
and the Hermite polynomials are
\begin{equation}
{\rm He}_3(x)=x^3-3x\,,\quad
{\rm He}_4(x)=x^4-6x^2+3\,, \quad 
{\rm He}_6=x^6-15x^4+45x^2-15\,.
\end{equation}

By considering those second order terms in this expansion we are able to recover the desired reduced skewness, $S_3$, with errors below $2$ percent.

\subsection{Theoretical and simulated matter PDF}
\label{ap:ap2}

In this appendix, we provide a comparison between the matter PDF predicted by \textsc{CosMomentum}, described in \autoref{sec:PDF_theory}, and the matter PDF calculated from UNIT dark matter particles. This test helps us identify possible sources of error in the intermediate steps of our modelling that may propagate to the description of our HI signal.

\begin{figure}[htbp]
    \centering
    \begin{subfigure}[t]{0.49\textwidth}
        \centering
        \includegraphics[width=\textwidth]{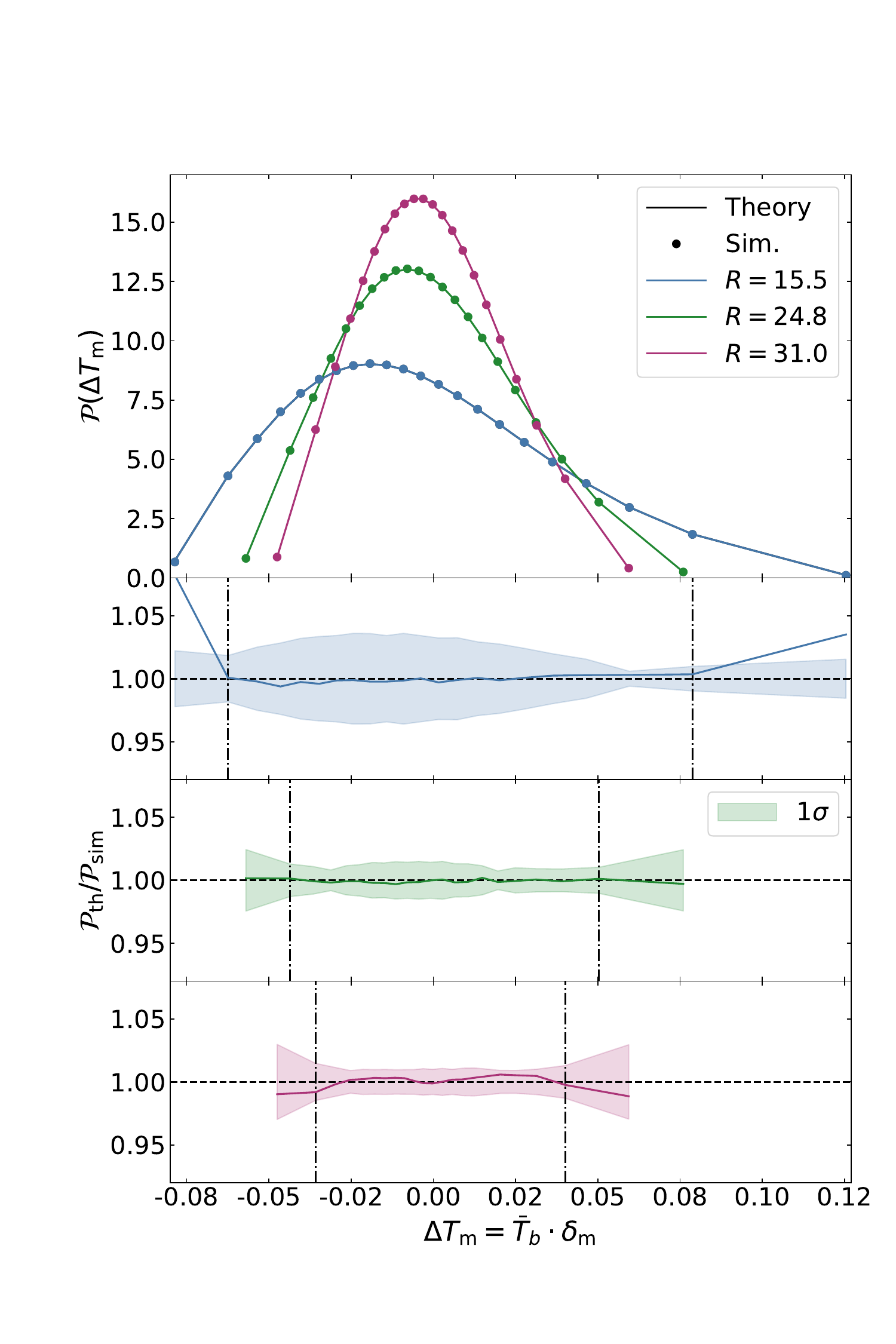}
        \caption{No observational systematics}
        \label{fig:PDF_CosMomentum_dm}
    \end{subfigure}
    \hfill
    \begin{subfigure}[t]{0.49\textwidth}
        \centering
        \includegraphics[width=\textwidth]{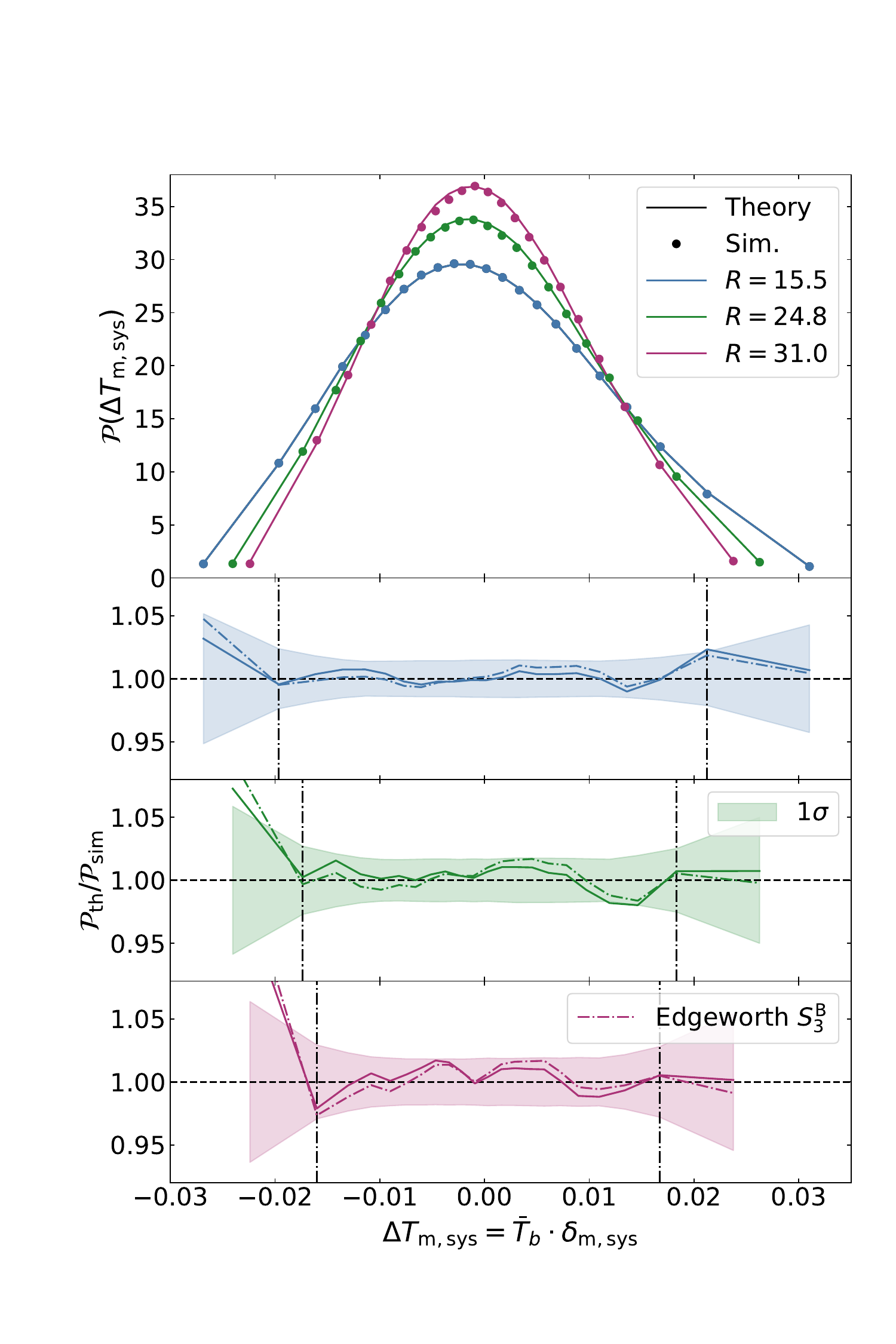}
        \caption{Considering observational systematics}
        \label{fig:PDF_CosMomentum_sys_dm}
    \end{subfigure}
    \caption{\textbf{Top:} Simulation-based DM PDFs with smoothing radii of $R = 15.5$, $24.8$, and $31.0$ ${\rm Mpc}/h$ from DM Intensity maps at $z = 1.321$ are shown as colored points, plotted as a function of the temperature fluctuations $\Delta T_{\rm m(,sys)}$. The corresponding PDFs obtained with \textsc{CosMomentum}, are shown as solid lines. \textbf{Bottom:} Ratio between the theoretical and simulated PDFs, with the jackknife-based $1\sigma$ uncertainty shown as shaded regions. Vertical black dashed lines indicate the scale cuts applied for each smoothing scale $R$ in the Fisher analysis (see section \ref{section: cosmological_parameter_estimation}), which exclude the first and two last quantiles. On the right, dot-dashed lines show the PDFs with an Edgeworth-based systematics correction to match the theoretical skewness, $S_3^{\rm B}$, computed in \autoref{sec:skewness}.}
    \label{fig:DM_PDF_CosMomentum_no_sys_and_sys}
\end{figure}

\autoref{fig:DM_PDF_CosMomentum_no_sys_and_sys} shows the simulated and theoretical matter PDFs, with and without systematics, for smoothing radii of $R = 15.5$, $24.8$, and $31.0$ ${\rm Mpc}/h$, together with their corresponding ratios, in the same format as \autoref{fig:PDF_CosMomentum_no_sys_and_sys}. In the case without systematics, we find good agreement between theory and simulations, with discrepancies well below $1\sigma$ over the range corresponding to the 5th--95th percentiles of the matter CDF (that is, without taking into account the first and last quantiles). Within this range, we do not identify any systematic source of error at this stage that would clearly propagate into our HI analysis.

For the case including systematics, we also find good agreement between theory and simulations, although the PDFs measured from the simulations are noticeably noisier. The skewness of the theoretical PDFs differs only very slightly from that of the simulated PDFs, but this difference is not visually apparent because it is masked by the additional noise in the simulations.

\subsection{Effective volume}
Beyond changes in the cumulants,  we also have to consider that the effective volume changes with the systematics kernels that act as additional smoothings. This is relevant for the estimation of the tracer stochasticity which uses an effective number count, and the overall level of noise. In particular in the case of the telescope beam that is a Gaussian angular smoothing which gives
\begin{equation}
\label{eq:Veff_beam}
V_{\rm eff}(R,R_{\rm beam})=\left[\int \frac{d^3k}{(2\pi)^3}
\,W_{\rm TH}^2(kR)\,
B^2(k_\perp)
\right]^{-1}
=\frac{\frac{4\pi}{3}  R^3}{1+\frac{3\sqrt{\pi }}{8}  \frac{R_{\rm beam}}{R} \left(\frac{R_{\rm beam}^2}{R^2}-2\right) \text{erf}\left(\frac{R}{R_{\rm beam}}\right)-\frac{3R_{\rm beam}^2}{4R^2} \exp\left(\frac{-R^2}{R_{\rm beam}^2}\right)}\,.
\end{equation}

\section{Testing Gaussianity of the data vector}
\label{ap:test_gaussianity}
To validate the Fisher analysis, which assumes a Gaussian likelihood, we perform a $\chi^2$ test to check if the bins of our statistics also follow a Gaussian distribution. We test all Jackknife realisations we considered in this work, $N_{\rm total} = 500$. We calculate the $\chi^2$ for each realisation as:

\begin{equation}
\label{eqn:chi_squared_gaussianity_checking}
    \chi^2 = \frac{1}{2} \left(S_\alpha - \langle S_\alpha \rangle \right) C_{\alpha\beta} \left(S_\beta - \langle S_\beta \rangle \right)
\end{equation}
    
We construct a probability distribution function for the $\chi^2$, $P(\chi^2)$, and we use a normal test to check if the distribution is sufficiently close to a Gaussian distribution. In particular we set $p > 0.05$ as a lower bound for the p-value.

\begin{figure}[htbp]
    \centering
    \begin{subfigure}[t]{0.49\textwidth}
        \centering
        \includegraphics[width=\textwidth]{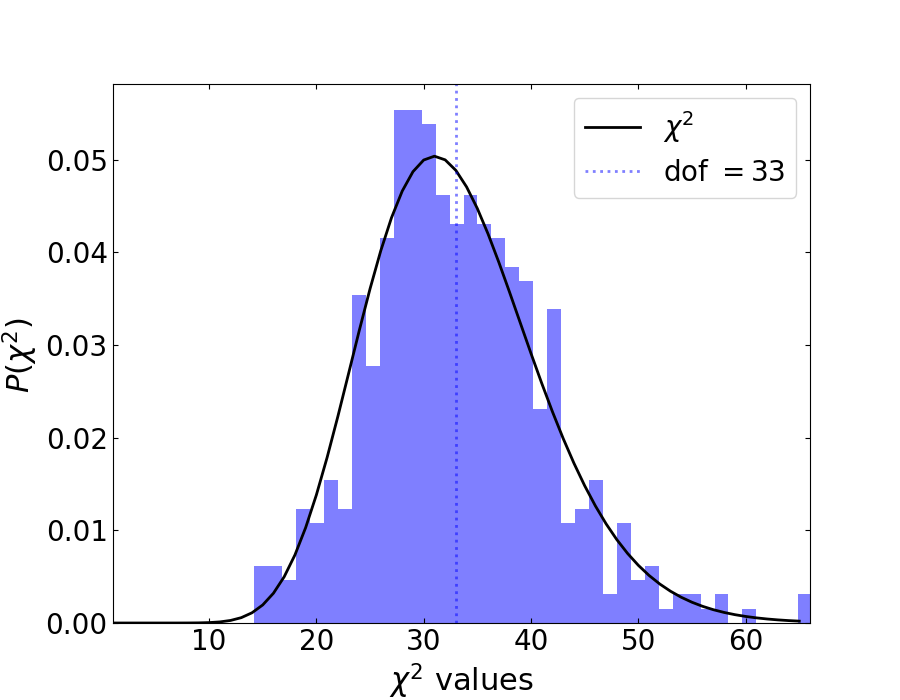}
        \caption{$S_\alpha = \{ P(k),\mathcal{P}(\Delta T_{\rm HI})$ (R = 31.0 ${\rm Mpc}/h$)\}}
        \label{fig:gaussianity_test_Pk_50}
    \end{subfigure}
    \hfill
    \begin{subfigure}[t]{0.49\textwidth}
        \centering
        \includegraphics[width=\textwidth]{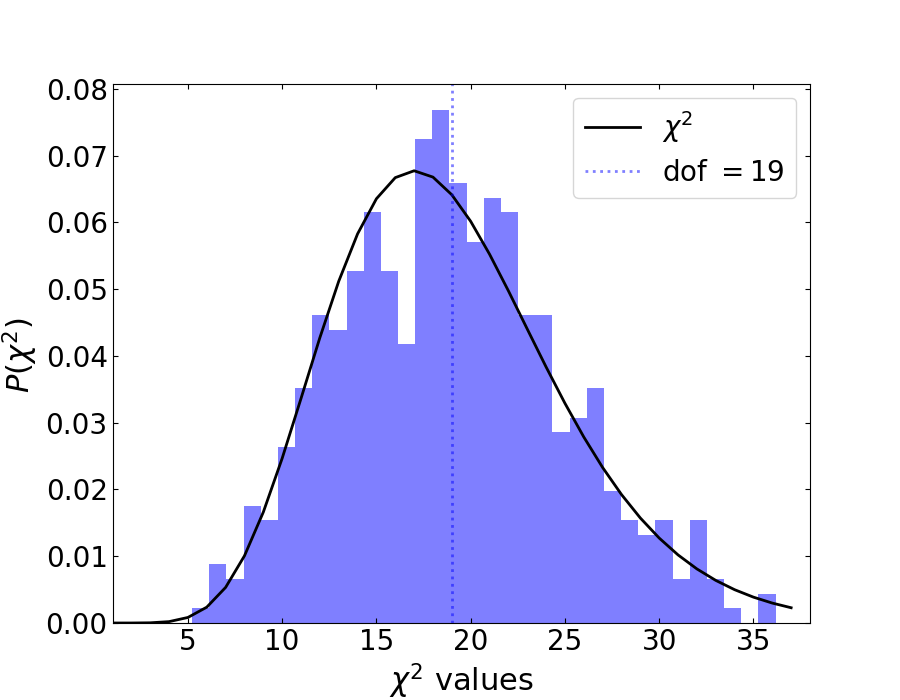}
        \caption{$S_\alpha = \{ \mathcal{P}(\Delta T_{\rm HI})$ (R = 31.0 ${\rm Mpc}/h$)\}}
        \label{fig:gaussianity_test_50}
    \end{subfigure}
    \caption{$\chi^2$ distribution from the data (blue histogram) for the PDF alone (right) and its combination with the power spectrum (left) compared to  a $\chi^2$ distribution with the appropriate number of degrees of freedom (dof) as expected for an underlying Gaussian distribution of data points (black solid line).}
    \label{fig:gaussianity_test}
\end{figure}

We checked results for the power spectrum and the PDFs, as well as for all combinations between the power spectrum and the PDFs. In \autoref{fig:gaussianity_test} we represent two examples: $\mathcal{P}(\Delta T_{\rm HI})$ (R = 31.0) ${\rm Mpc}/h$ and combining it with the power spectrum, $P(k)$. The distributions should be centered at $\rm{dof} = 19$ and $\rm{dof} = 33$, respectively, which are the number of degrees of freedom. Since we do not have that many realisations we expect to see noise in our distribution but on top of that, $P(\chi^2)$ passes the normal test with a p-value of $p = 0.2$, providing no evidence to reject the Gaussian hypothesis and remaining well within the expected $2\sigma$ range.

\section{HI Stochasticity}
\label{ap:HIstochasticity}

\begin{center}
    \begin{figure}[!t]
        \centering
    	\includegraphics[scale=0.65]{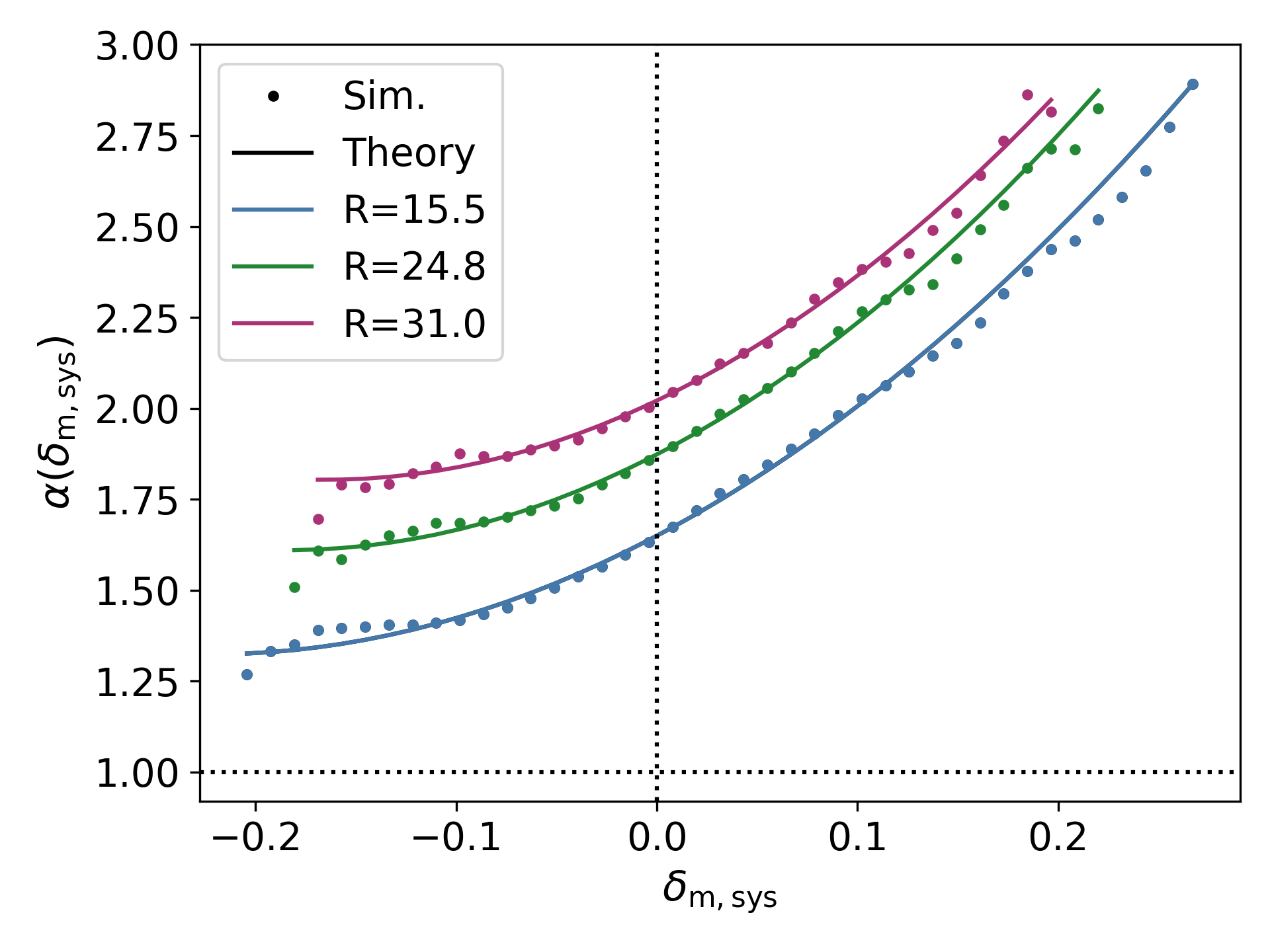}
    	\caption{UNIT-based shot noise as a function of dark matter overdensity with systematics for three different smoothing scales are shown as points. Quadratic best fits are shown as solid lines. The black dotted horizontal line, $\alpha = 1$, represents Poissonian shot noise.}
    \label{fig:stochasticity_vs_delta_withsys}    
    \end{figure}
\end{center}

In this appendix, we present the HI stochasticity for three different smoothing scales, including systematic effects in the analysis, as shown in \autoref{fig:stochasticity_vs_delta_withsys}. The points correspond to the UNIT-based measurements, while the solid lines show the best-fitting quadratic models (\autoref{eqn:alpha_HI}). The shot noise is systematically super-Poissonian and exhibits a clear quadratic increase with the dark matter overdensity, indicating that a quadratic parametrisation provides an adequate description of the measured stochasticity.

A similar trend is already present when systematic effects are not included. However, in this case the shot noise distribution measured from the simulations displays a more complex shape: the underdense tails are enhanced, whereas the overdense tails are suppressed relative to the quadratic model (see blue line in \autoref{fig:HI_DM_no_sys}).

\end{appendix}
\end{document}